\documentstyle[12pt]{article}
\makeatletter
\@addtoreset{equation}{section}
\makeatother
\renewcommand{\theequation}{\thesection.\arabic{equation}}
\begin{document}

\begin{center}
{\Large
An Axiomatic Approach to  Semiclassical  Perturbative  Gauge  Field
Theories
}
\\[1cm]
{{\large  O.Yu.Shvedov} \\[0.5cm]
{\it
Sub-Dept. of Quantum Statistics and Field Theory},\\
{\it Dept. of Physics, Moscow State University},\\
{\it 119992, Moscow, Vorobievy Gory, Russia}
}
\end{center}

\setcounter{page}{0}

\begin{flushright}
hep-th/0512352
\end{flushright}

\section*{Abstract}

Different approaches to axionatic field theory are  investigated.  The
main notions of semiclassical theory are the following:  semiclassical
states, Poincare   transformations,   semiclassical    action    form,
semiclassical gauge   equivalence  and  semiclassical  field.  If  the
manifestly covariant approach is used,  the  notion  of  semiclassical
state is  related to Schwinger sourse,  while the semicalssical action
is presented via the R-function of Lehmann,  Symanzik and  Zimmermann.
Semiclassical perturbation  theory  is constructed.  Its relation with
the S-matrix theory is investigated. Semiclassical electrodynamics and
non-Abelian gauge theories are studied, making us of the Gupta-Bleuler
and BRST approaches.

{\it Keywords:}
Maslov semiclassical   theory,   axiomatic   quantum   field   theory,
Bogoliubov S-matrix,  Lehmann-Symanzik-Zimmermann approach,  Schwinger
sources, gauge theories, BRST-quantization.

\footnotetext{e-mail: olegshv@mail.ru}

\footnotetext{This work was supported by the Russian  Foundation  for
Basic Research, project 05-01-00824.
}

\makeatletter
\@addtoreset{equation}{section}
\makeatother
\renewcommand{\theequation}{\thesection.\arabic{equation}}

\def\qp{
\mathrel{\mathop{\bf x}\limits^2},
\mathrel{\mathop{-i\frac{\partial}{\partial {\bf x}}}\limits^1} 
}
\def\gsim{{> \atop \sim}}
\def\lsim{{< \atop \sim}}
\def\simp{{\sim \atop {phys}}}
\def\ou#1{\underline{\overline{#1}}}

\def\lb#1{\label{#1}}
\def\l#1{\lb{#1}}
\def\r#1{(\ref{#1})}
\def\c#1{\cite{#1}}
\def\i#1{\bibitem{#1}}
\def\beq{\begin{equation}}
\def\eeq{\end{equation}}
\def\bez{\begin{displaymath}}
\def\eez{\end{displaymath}}
\def\beb#1\l#2\eeb{\begin{equation} \begin{array}{c} #1 \qquad
\end{array} \label#2  \end{equation}}
\def\bey#1\eey{\begin{displaymath}
\begin{array}{c} #1  \end{array}  \end{displaymath}}

\newpage

\section{Introduction}

A usual way to construct quantum field theory is as follows (see,  for
example, \c{1}).  First of all,  one  considers  the  classical  field
theory instead  of  the  quantum  one.  In order to make the classical
theory to be manifestly covariant,  one uses the  Lagrangian  approach
instead of  Hamiltonian.  Then  one  rewrites  a  manifestly covariant
Lagrangian theory to  the  Hamiltonian  language  and  postulates  the
"rules" of   canonical   quantization.   Qunatum   field  models  with
interaction are usually not  exactly  solvable;  therefore,  they  are
investigated with the help of formal perturbation theory.  Analogously
to quantum mechanical  case,  one  constructs  a  formal  perturbation
series for the S-matrix expressed via the T-exponent and for the Green
functions, a parameter of expansion is coupling constant.

An alternative way to construct quantum field theory  is  to  use  the
functional integral approach \c{2}:  quantum field Green functions are
written via the functional integrals (of  exponent  of  the  classical
action) analogously  to  quantum  mechanics  \c{2a}.  This approach is
formally manifestly covariant.  However,  one should take into account
that the  non-perturbative  definition  of a functional interal is not
rigorous.

Making use of the calculated Green functions,  one can reconstruct all
objects of  the  theory  with  the  help  of Whightmann reconstruction
theorem \c{3,4}.  The problem is that for this purpose it is necessary
to know the exact non-perturbative Green functions.

The main  difficulty  of  quantum  field  theory  is  the  problem  of
divergences. They arise in the functional integral approach,  as  well
as in  the  T-exponent  technique.  To eliminate the divergences,  one
should first  introduce  the  regularization.  It  may  be  manifestly
covariant (Pauli-Willars,  dimensional)  or non-covariant (ultraviolet
cutoff, dimensional). Then one should perform the renormalization.
Counterterms are added to the Lagrangian in order to  make  the  Green
functions finite.  When the non-covariant regularization is used,  one
should explicitly  check  Poincare  invariance.  For   theories   with
symmetries, it  is  also  necessary to check them in the renormalizaed
theory.

Another group of approaches to construct  quantum  field  perturbation
theory is based on the axiomatic approaches
\c{3,4,5,6,7,8,9}. One first  formulates  the  axioms  for  the  Green
functions and   related   objects.   This   allows  to  construct  the
renormalized theory without  renormalization.  Examples  of  axiomatic
theories are  Bogoliubov  S-matrix  theory \c{6} with switching on the
interaction, Schwinger          source          theory          \c{9},
Lehmann-Symanzik-Zimmermann (LSZ) approach \c{8} and S-matrix approach
of Bogoliubov,  Medvedev and Polivanov \c{6,7} which is equivalent  to
LSZ approach.

Making use of the Bogoliubov S-matrix theory, one can obtain
the renormalized  perturbation  theory.  One  uses  the  most  general
properties: Poincare  covariance,  unitarity  and causality.  However,
there is a non-uniqueness in axiomatic perturbation theory: each order
of the S-matrix is found up to a quasilocal operator.

There are  also  non-perturbative appproaches to quantum field theory.
One  of  them  is   a   semiclassical   approximation.   Examples   of
semiclassical   results   are   soliton  quantization  and  instantons
\c{10,10a}, quantum field theory in the external background \c{11} and
curved space-time \c{12}.  Semiclassical theory was developed,  making
use of the Hamiltonian and functional integral approaches.

The purpose of this paper  is  to  clarify  the  relationship  between
Hamiltonian   and   axiomatic  field  theories  in  the  semiclassical
approach. The semiclassical perturbation theory is constructed, making
use of the axiomatiic conceptions.

Section 2 deals with general properties of the semiclassical states of
the quantum field system.  The semiclassical analogs of the QFT axioms
are formulated. They should be satisfied for Hamiltonian and axiomatic
approaches.

Section 3 is devoted to  the  manifestly  covariant  approach  to  the
quantum field  theory.  Semiclassical  states are related to Schwinger
sources, while semiclassical action and fields are expressed  via  LSZ
R-functions.

The leading  order  of semiclassical expansion is developed in section
4. Semiclassical  perturbation  theory  is  developed  in  section  5.
Sections 6 and 7 are devoted to semiclassical gauge theories.

\section{General Structure of Semiclassical Perturbation Field Theory}

\subsection{Semiclassical States}

A usual way to develop semiclassical field theory (QFT in the external
background) is as follows.  The field $\varphi$ is presented as a  sum
of "classical", c-number part and fuctuation, "quantum" part, which is
small with respect to the classical one.  Action is  expanded  into  a
series in  powers  of  quantum  fluctuation.  In the leading order the
quadratic part of action only is taken to account.

Semiclassical methods can be applied iff the Lagrangian of the theory
$\cal L$ depends on the small parameter
$h$ ("Planck constant") as follows (the scalar case is considered  for
the simplictiy):
\beq
{\cal L} = \frac{1}{2} \partial_{\mu} \varphi \partial^{\mu} \varphi -
\frac{1}{h} V(\sqrt{h} \varphi).
\l{2.1}
\eeq
with $V(\Phi)$ being a scalar potential.
The classical c-number component is of the order
$1/\sqrt{h}$.

In terms of quantum states and equation of motion,  this semiclassical
method can be reformulated as follows.  The  following  time-dependent
state vector \c{13,13a} is considered:
\beq
\Psi(t) \simeq e^{\frac{i}{h}S(t)}
e^{\frac{i}{\sqrt{h}} \int d{\bf x}
[\Pi({\bf x},t)  \hat{\varphi}({\bf  x})  - \Phi({\bf x},t) \hat{\pi}({\bf
x})]} f(t).
\l{2.2}
\eeq
Here $S(t)$ is a real c-number finction of $t$,
$\Phi({\bf
x},t)$ and  $\Pi({\bf  x},t)$  are  classical  fields  and  canonucally
conjugated momenta,
$\hat{\varphi}({\bf x})$  and  $\hat{\pi}({\bf  x})$ are quantum field
and momentum operators, $f(t)$ is a regular as $h\to 0$ state vector.
Substitute vector  \r{2.2}  to  Schrodinder  equation  for  the theory
\r{2.1}. One obtains in  the  leading  order  classical  equations  of
motion for $\Pi({\bf x},t)$,  $\Phi({\bf x},t)$, formula for classical
action for $S(t)$ and equation with quadratic Hamiltonian for  $f(t)$.
This is in agreement with method of extracting a c-number component of
the field.

In terms of the semiclassical Maslov theory
\c{14,15,16}, state
\r{2.2} corresponds to the Maslov complex germ in a point.
There is also a theory of Lagrangian manifolds with complex germs.  It
can be also generalized to quantum field case. One considers
superpositions of states
\r{2.2} \c{17}:
\beq
\Psi(t) \simeq const \int d\alpha e^{\frac{i}{h}S(\alpha,t)}
e^{\frac{i}{\sqrt{h}} \int d{\bf x}
[\Pi({\bf x},\alpha,t)  \hat{\varphi}({\bf  x})
- \Phi({\bf x},\alpha,t) \hat{\pi}({\bf
x})]} f(\alpha,t).
\l{2.4a}
\eeq
where $\alpha$ is an additional
$k$-dimensional parameter.

Semiclassical states
\r{2.4a} cannot  be  considered within the framework of extracting the
c-number component.  On the other hand,  such semiclassical states are
very important.  When systems with integrals of motion are considered,
it is  necessary  to  construct  the   state   satisfying   additional
conditions. For example, when one quantize kink solution
\c{10}, quantum state should be an eigenstate of  energy  and  momentum
operators. Projection on the eigenspace of the momentum operator leads
to the superposition
\r{2.4a}. When   one   uses  the  simpler  semiclassical  methods  and
investigates states of  the  form  \r{2.2}  only,  one  comes  to  the
well-known difficulty of soliton zero modes \c{10}.

Semiclassical state   vectors   \r{2.2}  and  \r{2.4a}  possesses  the
following geometric  interpretation  \c{18}.   denote   the   set   of
classical variables
$(S,\Pi({\bf  x},\Phi({\bf  x}))$ as
$X$,
the operator that maps vector
$f$ to the state
\r{2.2} is denoted as
$K_X^h$ ("the canonical operator"):
\beq
K^h_X f \equiv
e^{\frac{i}{h}S}
e^{\frac{i}{\sqrt{h}} \int d{\bf x}
[\Pi({\bf x})  \hat{\varphi}({\bf  x})  - \Phi({\bf x}) \hat{\pi}({\bf
x})]} f.
\l{2.2a}
\eeq
Then the semiclassical state \r{2.2a} can be viewed as a point on  the
space of  a  vector  bundle ("semiclassical bundle").  The base of the
bundle is set $\{X\}$ of classical states, fibres
$\{f\}$  are state spaces in the external background
$X$.   States of the more general form
\r{2.4} are written as
\beq
\int d\alpha     K^h_{X(\alpha)}    f(\alpha),    \quad    \alpha    =
(\alpha_1,...,\alpha_k);
\l{2.4}
\eeq
they can be identified with
$k$-dimensional surfaces on the semiclassical bundle.

Presentation of  semiclassical  form  in  the  form  \r{2.2a}  is  not
manifestly covariant. There are space and time coordinates. It happens
that the manifestly covariant  form  of  the  state  \r{2.2a}  is  the
following:
\beq
\Psi \simeq
e^{\frac{i}{h}\overline{S}}
Texp\{\frac{i}{\sqrt{h}} \int dx J(x) \hat{\varphi}_h(x)\} \overline{f}
\equiv
e^{\frac{i}{h}\overline{S}}
T_J^h \overline{f}.
\l{2.3}
\eeq
Here
$\overline{S}$ is a real number,  $J(x)$ is a real function (classical
Schwinger   source),   $\hat{\varphi}_h(x)$   is  a  Heisenberg  field
operator,  $\overline{f}$ is a state vector being regular as $h\to 0$.
The  Schwinger  source  $J(x)$  should be rapidly damping at space and
time infinity.

One can find that state \r{2.3} approximately coincides as
$h\to  0$  with \r{2.2}, and find a correspondence between
$(S,\Pi,\Phi,f)$ and $(\overline{S},J,\overline{f})$.

In the covariant approach,  the base of the semiclassical bundle is  a
set  of $\{X = (\overline{S},J)\}$,  the operator $K^h_X$ has the form
$e^{\frac{i}{h}\overline{S}} T_J^h$.  Superpositions of states \r{2.3}
are written as \r{2.4}.

\subsection{Properties of semiclassical states}

The main  principles  of  the axiomatic quantum field theory are \c{3}
Poincare invariance,  unitarity and causality.  The general structures
are introduced: the quantum Poincare transformation operator
${\cal U}_g^h$ corresponding to the classical Poincare transformation
$g\equiv  (a,\Lambda)$ of the form
$x'{}^{\mu} = \Lambda^{\mu}_{\nu} x^{\nu} + a^{\mu}$.
The group property should be satisfied:
\beq
{\cal U}_{g_1g_2}^h = {\cal U}_{g_1} {\cal U}_{g_2}.
\l{2.6a}
\eeq
The Heisenberg field operator
$\hat{\varphi}_h(x)$ should   satisfy   the   property   of   Poincare
invariance:
\beq
{\cal U}_{g^{-1}}^h     \hat{\varphi}_h(x)      {\cal      U}_g^h      =
\hat{\varphi}_h(w_gx), \qquad w_gx = \Lambda^{-1}(x-a)
\l{2.6w}
\eeq
Formula
\r{2.6w} is written for the scalar case;
for the more complicated cases it should be modified.

Analogous structures should arise for the semiclassical case as well.

The following commutation relations between  quantum  field  operators
and canonical operator $K^h_X$ are satisfied:
\beb
{\cal U}^h_g K_X^h f = K^h_{u_gX} \underline{U}_g(u_gX \gets X) f;\\
\sqrt{h} \hat{\varphi}_h(x) K_X^h f = K_X^h \underline{\Phi}(x|X) f,
\l{2.5}
\eeb
Here $u_gX$ is a Poincare transformation of the classical state,
$\underline{U}_g(u_gX \gets X)$ is an unitary operator taking  initial
state in the external background
$X$  to final state in the external background
$u_gX$. The operator
$\underline{U}_g(u_gX \gets X)$ is supposed  to  be  expanded  into  an
asymptotic series in $\sqrt{h}$.
Therefore, in the  semiclassical  theory  the  Poincare  group  is  an
automorphism group of the semiclassical bundle. Semiclassical field
$\underline{\Phi}(x|X)$ (distribution) is  also  viewed  as  a  formal
asymptotic series in $\sqrt{h}$. In the leading order, it has a
$c$-number form
$\Phi(x|X)$ (it is a classical field):
\beq
\underline{\Phi}(x|X) = \Phi(x|X) + \sqrt{h} \Phi^{(1)}(x|X) + ...
\l{2.6}
\eeq
It follows from the group property
\r{2.6a} and commutation relation
\r{2.5} that
\beb
u_{g_1g_2} = u_{g_1}u_{g_2};\\
\underline{U}_{g_1g_2}(u_{g_1g_2}X \gets X) =
\underline{U}_{g_1}(u_{g_1g_2}X \gets u_{g_2}X) =
\underline{U}_{g_2}(u_{g_2}X \gets X);
\l{2.6z}
\eeb
Making use of \r{2.6w}, one finds:
\beq
\underline{\Phi}(x|u_gX) \underline{U}_g(u_gX \gets X) =
\underline{U}_g(u_gX \gets X) \underline{\Phi}(w_gx|X).
\l{2.6v}
\eeq

There are also the  following additional
structures  of  semiclassical  field  theory.
To calculate the norm of the semiclassical state,  one should evaluate
the integral of the form:
\beq
\int d\alpha  d\alpha'
(K^h_{X(\alpha)}  f(\alpha),  K^h_{X(\alpha')} f(\alpha'))
\l{2.6c}
\eeq
If the difference between classical configurations
$X(\alpha)$ and $X(\alpha')$
is of the order $O(1)$ as
$h\to 0$,  the inner product
$(K^h_{X(\alpha)}  f(\alpha),  K^h_{X(\alpha')} f(\alpha'))$
is exponentially   small.   This   means   that   the   state  vectors
corresponding to different  classical  configurations  are  orthogonal
with exponential accuracy. Therefore, the integrand in
\r{2.6c} is not exponentially small for the caase of small
$\alpha - \alpha'$ only; more preciesely, for
$\alpha - \alpha' \sim \sqrt{h}$.

Therefore, it is necessary to expand the states
$K^h_{X+h\delta X} f$ and
$K^h_{X+\sqrt{h}\delta X} f$
into series in
$\sqrt{h}$ as $h\to 0$.
It happens that in the leading order in $h$
\beq
K^h_{X+h\delta X} f \simeq
e^{-i\omega_X[\delta X]} K^h_Xf,
\l{2.7a}
\eeq
expression
$\omega_X[\delta X]$ is an action 1-form on classical phase space.
If we consider the case
$\omega_X[\delta X] = 0$, the shift of classical variable of the order
$\sqrt{h}$ leads to the relation
\beq
K^h_{X+\sqrt{h}\delta X} f \simeq
K^h_X e^{-i\omega_X^{(1)}[\delta X]} f,
\qquad
\omega_X[\delta X] = 0
\l{2.7b}
\eeq
with an operator-valued
1-form $\omega_X^{(1)}[\delta X]$.

To write down all the orders of the expansion in
$\sqrt{h}$,  it is convenient to
start from the commutation rule between differentiation  operator  and
canonical operator:
\beq
ih\frac{\partial}{\partial\alpha_a} K^h_{X(\alpha)}  = K^h_{X(\alpha)}
\underline{\omega}_{X(\alpha)} [\frac{\partial X}{\partial \alpha_a}].
\l{2.7}
\eeq
Making use of formula
\r{2.7}, one  introduces  to  the  semiclassical   mechanics   a   new
structure,
operator-valued 1-form
$\underline{\omega}_X[\delta    X]$. It maps a tangent vector
$\delta X$ for the base of the semiclassical bundle to the operator
$\underline{\omega}_X[\delta X]$.
It is expanded into an asymptotic series in $\sqrt{h}$. In the leading
order, it is a c-number:
\beq
\underline{\omega}_X[\delta X]  =  \omega_X[\delta   X]   +   \sqrt{h}
\omega_X^{(1)}[\delta X] + ...
\l{2.8b}
\eeq
Rules
\r{2.7a}  and \r{2.7b}  are obtained as follows
\c{19}.
One expands the operator $K^h_{X(\alpha + \sqrt{h}\beta)}$  in  $\sqrt{h}$
as
\beq
K^h_{X(\alpha  +   \sqrt{h}\beta)} = K^h_{X(\alpha)} V_h(\alpha,\beta);
\l{2.9}
\eeq
then one differentiates the relation \r{2.9} with respect to $\beta_a$
and  obtains  an  equation  for  $V_h$.  Its solution gives a relation
\r{2.7b}.

Consider the substitution $\alpha' = \alpha + \sqrt{h} \beta$ for  eq.
\r{2.6c}.  According to eqs.\r{2.7a} and \r{2.7b},  the integrand will
be rapidly oscillating,  except for the case of the  Maslov  isotropic
condition:
\beq
\omega_{X(\alpha)} [\frac{\partial X}{\partial \alpha_a}] = 0.
\l{2.12}
\eeq
The superposition \r{2.4} can be investigated under condition \r{2.12}
only. Otherwise,  the  exponentially  small  norm of the state \r{2.4}
will be larger than the accuracy.

Investigate the properties of the new object $\underline{\omega}$.

First, notice that for multidimensional
$\alpha= (\alpha_1,...,\alpha_k)$ it is possible to write
$[ih\frac{\partial}{\partial \alpha_a},
ih\frac{\partial}{\partial \alpha_b}] = 0$;
commuting this relation with the canonical operator
$K^h_X$, one finds:
\beq
\left[
\omega_{X(\alpha)} [\frac{\partial X}{\partial \alpha_a}];
\omega_{X(\alpha)} [\frac{\partial X}{\partial \alpha_b}]
\right] = - ih
\left[
\frac{\partial}{\partial \alpha_a}      \underline{\omega}_{X(\alpha)}
[\frac{\partial X}{\partial \alpha_b}]
-
\frac{\partial}{\partial \alpha_b}      \underline{\omega}_{X(\alpha)}
[\frac{\partial X}{\partial \alpha_a}]
\right],
\l{2.8}
\eeq
or, in terms of differential forms,
\beq
[\underline{\omega}_X[\delta X_1],\underline{\omega}_X[\delta X_2]] =
- ih d\underline{\omega}_X(\delta X_1,\delta X_2).
\l{2.8a}
\eeq
Furthermore, it follows from the properties
\bez
[ih\frac{\partial}{\partial \alpha_a},\hat{\varphi}_h(x)] = 0,
\qquad
[ih\frac{\partial}{\partial \alpha_a},{\cal U}_g^h] = 0
\eez
that
\beq
\underline{U}_g(u_gX \gets    X)
\underline{\omega}_X[\frac{\partial X}{\partial \alpha_a}] =
\underline{\omega}_{u_gX}[\frac{\partial (u_gX)}{\partial \alpha_a}]
\underline{U}_g(u_gX \gets X) +
ih \frac{\partial}{\partial \alpha_a} \underline{U}_g(u_gX\gets X);
\l{2.13m}
\eeq
\beq
ih\frac{\partial}{\partial \alpha_a} \underline{\Phi}(x|X) =
\left[\underline{\Phi}(x|X);
\underline{\omega}_X[\frac{\partial X}{\partial \alpha_a}]\right].
\l{2.13n}
\eeq
Relation
\r{2.13m} is useful for investigating the properties of the  evolution
operator.

\subsection{Equivalence of semiclassical states}

A remarkable  feature  of  the covariant approach to the semiclassical
field theory is the following.  Starting from the classical  Schwinger
source $J$,  one  uniquely  reconstructs  classical  field and momenta
entering to eq.\r{2.2a}. On the other hand, this correspondence is not
one-to-one. Two   different   sources   may  correspond  to  the  same
configuration
$(\Pi({\bf  x}),\Phi({\bf x}))$.
Therefore, there is an equivalence relation  on  the  classical  space
(base of the semiclassical bundle). Moreover, it happens that for each
pair of equivalent classical states
$X_1 \sim X_2$ semiclassical states
\beq
K^h_{X_1} \overline{f}_1 \simeq
K^h_{X_2} \overline{f}_2
\l{2.13a}
\eeq
approximately coincide under a certain condition between
$\overline{f}_1$ and $\overline{f}_2$:
\bez
\overline{f}_2 = \underline{V}(X_2\gets X_1) \overline{f}_1.
\eez
Here
$\underline{V}(X_2\gets X_1)$ is an unitary operator.
Thus, an equivalence relation arises on the semiclassical bundle.

Investigate the properties of the operator
$\underline{V}(X_2\gets X_1)$.  Notice that the following relation
\beq
\underline{V}(X_3\gets X_1) =
\underline{V}(X_3\gets X_2)
\underline{V}(X_2\gets X_1)
\l{2.14}
\eeq
is satisifed. It follows from the properties
\bez
{\cal U}_g^h K^h_{X_1} \overline{f}_1 \simeq
{\cal U}_g^h K^h_{X_2} \overline{f}_2;
\quad
\sqrt{h} \hat{\varphi}_h(x) K^h_{X_1} \overline{f}_1 \simeq
\sqrt{h} \hat{\varphi}_h(x) K^h_{X_2} \overline{f}_2
\eez
that
\beq
\underline{V}(u_gX_2 \gets u_gX_1) \underline{U}_g(u_gX_1 \gets X_1) =
\underline{U}_g(u_gX_2 \gets X_2)
\underline{V}(X_2 \gets X_1).
\l{2.15}
\eeq
\beq
\underline{\Phi}(x|X_2) \underline{V}(X_2 \gets X_1) =
\underline{V}(X_2 \gets X_1) \underline{\Phi}(x|X_1).
\l{2.16}
\eeq
Finally, let
$(X_1,\overline{f}_1)$ and   $(X_2,\overline{f}_2)$   depend   on  the
parameter
$\alpha$. Then one writes
\bez
ih \frac{\partial}{\partial \alpha_a} K^h_{X_1} \overline{f}_1 \simeq
ih \frac{\partial}{\partial \alpha_a}K^h_{X_2} \overline{f}_2.
\eez
Therefore,
\beb
\underline{V}(X_2 \gets X_1)
\underline{\omega}_{X_1} [\frac{\partial X_1}{\partial \alpha_a}] =
\\
\underline{\omega}_{X_2} [\frac{\partial X_2}{\partial \alpha_a}]
\underline{V}(X_2 \gets X_1)
+ ih \frac{\partial}{\partial \alpha_a}
\underline{V}(X_2 \gets X_1)
\l{2.17}
\eeb
as $X_1(\alpha) \sim X_2(\alpha)$.

\subsection{System of axioms of the semiclassical field theory}

Thus, all  the  problems  of  semiclassical field theory can be solved
within the perturbation framework under certain  conditions  (axioms).
They should   be   satisfied   for   both  Hamiltonian  and  covariant
approaches. They will be denoted as G1, G2, G3, G4, G5.

{\bf G1}.  {\it  A semiclassical bundle is given.
Space of the bundle is  set of all semiclassical states.
The base
${\cal X} = \{X\}$ is set of classical states
$X$,
fibres
${\cal F}_X$ are
Hilbert spaces of quantum states in the external background
$X$.
}

{\bf G2.}  {\it  The  Poincare  group  is  presented  as  a  group  of
automorphisms of the semiclassical bundle.  Classical  transformations
$u_g:    {\cal    X}    \to    {\cal   X}$   and   unitary   operators
$\underline{U}_g(u_gX  \gets  X):  {\cal  F}_X  \to  {\cal  F}_{u_gX}$
expanded   in  $\sqrt{h}$  are  given.  The  properties  \r{2.6z}  are
satisfied. }

{\bf G3.} {\it An operator-valued
1-форма
$\underline{\omega}$ on the semiclassical bundle is given.
It maps tangent vector
$\delta X$ to Hermitian operator
$\underline{\omega}_X[\delta X]$. It is expanded in
$\sqrt{h}$ according to eq.
\r{2.8b} ($\omega_X[\delta X]$ is a c-number) and satisfied properties
\r{2.8} and \r{2.13m}.
}

{\bf G4.} {\it
An equivalence relation may be introduced on the semiclassical bundle.
For  each  pair  $X_1  \sim  X_2$  of equivalent points of the base an
unitary  opertator  $\underline{V}(X_2\gets  X_1)$  is  given.  It  is
expanded in $\sqrt{h}$.  Relations \r{2.14}, \r{2.15} and \r{2.17} are
satisfied. }

{\bf G5.} {\it
An operator-valued distribution
$\underline{\Phi}(x|X)$  ("semiclassical field") is given.
It is expanded in
$\sqrt{h}$  (eq.\r{2.6}, $\Phi(x|X)$ is a
c-number)  and satisfies the relations
\r{2.6v}, \r{2.13n} and \r{2.16}.
}

\section{Specific features of the covariant approach to the semiclassical field
theory}

\subsection{
Objects of the semiclassical theory and LSZ R-functions
}

Let us investigate  the  objects  of  the  covariant  formulations  of
semiclassical field theory.

Semiclassical states in the covariant approach are presented in a form
\r{2.3}.  Therefore,  the base of the semiclassical bundle consists of
the  classical  states $X = \{\overline{S},J\}$,  while all the spaces
${\cal F}_X$ coincide.

It follows from eq.\r{2.6w} that
\bez
{\cal U}^h_g   T^h_J   \overline{f}   =   T^h_{u_gJ}   {\cal   U}^h_g
\overline{f},
\eez
where $u_gJ(x) = J(w_gx)$,  $w_g$  has the form
\r{2.6w}.  Therefore, transformation
$u_g$ is known explicitly, satisfy property
$u_{g_1g_2}
= u_{g_1} u_{g_2}$, the operator
$\underline{U}_g(u_gX \gets  X)  \equiv
\underline{U}_g =  {\cal  U}^h_g$
does not depend on
$X$ and satisfies the group property and field covariance relation:
\beq
\underline{U}_{g_1g_2} = \underline{U}_{g_1} \underline{U}_{g_2},
\quad
\underline{U}_{g^{-1}} \hat{\varphi}_h(x)       \underline{U}_g        =
\hat{\varphi}_h(w_gx).
\l{3.1a}
\eeq

It happens that in the covariant approach the 1-form
$\underline{\omega}$ is related to another important object, the
LSZ R-function
\c{8,3} of the form
\beq
\underline{\Phi}_R(x|J)
\equiv - ih (T^h_J)^+ \frac{\delta T^h_J}{\delta J(x)}.
\l{3.1}
\eeq
To check this statement, notice that the differentiation (or variation)
operator commutes with the operator
$K^h_{\overline{S},J} \equiv  e^{\frac{i}{h}\overline{S}} T^h_J$ as
\bez
ih \delta K^h_{\overline{S},J} = K^h_{\overline{S},J}
[-\delta \overline{S} - \int dx \underline{\Phi}_R(x|J) \delta J(x)];
\eez
therefore,
\beq
\underline{\omega}_X[\delta X] =  -  \delta  \overline{S}  -  \int  dx
\underline{\Phi}_R(x|J) \delta J(x),
\l{3.2}
\eeq
Here   $\underline{\Phi}_R(x|J)$ and  $\underline{\omega}$,
are expanded in
$\sqrt{h}$; one writes
\beq
\underline{\Phi}_R(x|J) =
\Phi_R(x|J) + \sqrt{h} \Phi_R^{(1)}(x|J) + ...
\l{3.1aa}
\eeq
The c-number function
$\Phi_R(x|J)$ is called as a retarded classical field generated by the
Schwinger source $J$.

It is shown in
\c{20} that for the model
\r{2.1}  $\Phi_R(x|J)$ is a solution of the equation
\beq
\partial_{\mu} \partial^{\mu} \Phi_R(x|J)
+ V'(\Phi_R(x|J)) = J(x),
\quad
\Phi_R\vert_{x\lsim supp J} = 0.
\l{3.2a}
\eeq
which vanishes as $x^0 \to -\infty$.

The following properties of LSZ R-functions  are  well-known  \c{8,3}.
They are corollaries of \r{3.1}.

1. The Hermitian property
\beq
\underline{\Phi}_R^+(x|J) = \underline{\Phi}_R(x|J).
\l{3.3o}
\eeq

2. The Poincare invariance property
\beq
\underline{U}_{g^{-1}}
\underline{\Phi}_R(x|u_gJ) \underline{U}_g =
\underline{\Phi}_R(w_gx|J).
\l{3.3a}
\eeq

3. The Bogoliubov causality property: R-function
$\underline{\Phi}_R(x|J)$
depends only on the source $J$ at the preceeding time moments.
Making use of the standard notations
$x>y$ iff $x^0-y^0 \ge |{\bf x} - {\bf y}|$,
$x<y$ iff $x^0-y^0 \le |{\bf x} - {\bf y}|$,
$x\sim y$ iff $|x^0-y^0| < |{\bf x} - {\bf y}|$, one rewrites
the Bogoliubov condition as
\beq
\frac{\delta \underline{\Phi}_R (x|J)}{\delta J(y)} = 0,
\quad y \gsim x.
\l{3.3b}
\eeq

4. Commutation relation
\beq
[\underline{\Phi}_R(x|J);\underline{\Phi}_R(y|J)] = -ih \left(
\frac{\delta \underline{\Phi}_R(x|J)}{\delta J(y)} -
\frac{\delta \underline{\Phi}_R(y|J)}{\delta J(x)} \right).
\l{3.3c}
\eeq

5. Boundary condition at
$-\infty$.  If
$x\lsim y$ for all $y\in supp J$,  the LSZ R-functiion does not depend
on the source:
\beq
\underline{\Phi}_R(x|J) = \hat{\varphi}_h(x) \sqrt{h},
\quad
x \lsim supp J.
\l{3.3d}
\eeq
In particular, the classical retarded field vanishes as
$x^0 \to -\infty$.

The semiclassical field operator
$\underline{\Phi}(x|J)$    can be also expressed via the
R-function. Namely, at
$+\infty$ the following property is satisfied:
\beq
\underline{\Phi}(x|J) = \underline{\Phi}_R(x|J), \quad
x \gsim supp J.
\l{3.4}
\eeq
In particular,  the  classical  field $\Phi(x|J)$ corresponding to the
source $J$ coincide with the classical retarded field $\Phi_R(x|J)$ at
$x\gsim  supp  J$.  The  case when the property $x\gsim supp J$ is not
satisfied is considered below.

\subsection{
Equivalence of semiclassical states
}

Investigate the property of equivalence of semiclassical states. It is
convenoent to  consider  first the partial case when the semiclassical
state is equivalent to the state
$J=0$. We say that
$J\sim 0$ iff
\beq
T^h_J \overline{f}      \simeq      e^{\frac{i}{h}     \overline{I}_J}
\underline{W}_J \overline{f}
\l{3.5}
\eeq
for some  number  $\overline{I}_J$  and   operator   $\underline{W}_J$
presented as a formal asymptotic series.

It is shown in \c{20} for the model \r{2.1} that  the  source  $J$  is
equivalent to zero iff the retarded field generated by $J$ vanishes at
$+\infty$.

Analogously to   \c{20},   one   derives the  following
properties.

1. Poincare invariance.
\beq
\underline{U}_g \underline{W}_J        \underline{U}_{g^{-1}}        =
\underline{W}_{u_gJ}, \qquad \overline{I}_{u_gJ} = \overline{I}_J;
\l{3.6}
\eeq

2. Unitarity
\beq
\underline{W}_J^+ =
\underline{W}_J^{-1};
\l{3.7}
\eeq

3. Bogoliubov causality:
as $J+\Delta J_2 \sim 0$,  $J+\Delta
J_1 + \Delta J_2 \sim 0$ and $supp \Delta J_2  \gsim  supp  \Delta  J_1$
the operator
$(\underline{W}_{J+\Delta J_2})^+ \underline{W}_{J+\Delta J_1
+ \Delta   J_2}$   and number $-   \overline{I}_{J+\Delta   J_2}    +
\overline{I}_{J+\Delta J_1 + \Delta J_2}$ do not depend on
$\Delta J_2$.

4. Variational property:
\beq
\delta \overline{I}_J - ih \underline{W}_J^+ \delta \underline{W}_J =
\int dx \underline{\Phi}_R(x|J) \delta J(x),
\l{3.8}
\eeq
which is valid as
$J\sim 0$ and $J+\delta J \sim 0$.

5. Boundary condition at $+\infty$:
\beq
\underline{\Phi}_R(x|J) =       \underline{W}_J^+       \hat{\varphi}(x)       \sqrt{h}
\underline{W}_{J}, \quad x \gsim supp J.
\l{3.9}
\eeq

It follows from
\r{3.9} that the retarded classical field generated by the source
$J\sim  0$ will vanish at
$+\infty$.
For the model
\r{2.1},  an inverse statement is also valid:
{\it   for any field configuration $\Phi_c(x)$ with the compact support
one can uniquely choose a source
$J\sim  0$
(denoted as $J = J_{\Phi_c} = J(x|\Phi_c)$;
for example \r{2.1}, it is found from the relation
\r{3.2a})
generating
$\Phi_c(x)$ as a retarded classical field:
$\Phi_c(x) = \Phi_R(x|J)$;
it satisfies the locality condition
$\frac{\delta J(x|\Phi_c)}{\delta \Phi_c(y)} = 0$ as $x\ne y$.
}

It is possible to  treat  this  statement  as  a  basic  postulate  of
semiclassical field  theory.  Then the theory may be developed without
additional postulating  classical  stationary  action  principle   and
canonical commutation relation.

Namely, it  follows  from  eq.\r{3.8} in the leading order in $h$ that
the functional
\beq
I[\Phi_c] = \overline{I}_{J_{\Phi_c}} -
\int dx J_{\Phi_c}(x) \Phi_c(x)
\l{3.10}
\eeq
satisfies the "classical equation of motion"
\beq
J_{\Phi_c}(x) = - \frac{\delta I[\Phi_c]}{\delta \Phi_c(x)}.
\l{3.11}
\eeq
The functional $I[\Phi]$ should satisfy the locality condition
\beq
\frac{\delta^2I}{\delta \Phi_c(x) \delta \Phi_c(y)} = 0.
\quad x\ne y
\l{3.11a}
\eeq
This means that it is presented as an integral of a local Lagrangian.

Relation \r{3.11}  allows  us  to  reconstruct  the classical retarded
field,  making use of known  source  $J\sim  0$,  since  the  boundary
condition at  $-\infty$  is  known.  It  follows  from  the Bogoliubov
causality condition that the retarded field depends only on
$J$ at  the  preceeding  time  moments.  If  the  sourse  $J(x)$  is  not
equivalent to zero, it can be modified at $+\infty$ and transformed to
the sourse equivalent to zero.
Therefore, the relation
\beq
\frac{\delta I}{\delta \Phi_c} [\Phi_R(\cdot|J)] = - J(x),
\quad
\Phi_R\vert_{x<supp J} = 0
\l{3.11b}
\eeq
is valid for all sourses
$J$.
For the case $x\gsim supp J$,
the property
\r{3.11b}
is taken to the classical field equation
\beq
\frac{\delta I}{\delta \Phi_c(x)} [\Phi(\cdot|J)] = 0.
\l{3.11c}
\eeq
This is  a  classical  stationary  action  principle.  It  is viewed a
coroollory of other general principles of semiclassical field theory.

Thus, we see that classical action  $I[\Phi_c]$  in  field  theory  is
related  with the phase of the state $T_J^h \overline{f}$ as $J\sim 0$
according to eq.\r{3.10}.

Let us rewrite the properties of the  operator  $\underline{W}_J$  via
the    field    $\Phi_c$.    Denote    $\underline{W}[\Phi_c]   \equiv
\underline{W}_{J_{\Phi_c}}$.

1. Poincare invariance.
\beq
\underline{U}_g \underline{W}[{\Phi}_c]  \underline{U}_{g^{-1}}
= \underline{W}[u_g{\Phi}_c].
\l{3.12a}
\eeq

2. Unitarity.
\beq
\underline{W}^+[{\Phi}_c] = (\underline{W}[{\Phi}_c] )^{-1}.
\l{3.12b}
\eeq

3. Bogoliubov causality.
\beq
\frac{\delta}{\delta {\Phi}_c(y)}
\left(
\underline{W}^+[{\Phi}_c]
\frac{\delta \underline{W}[{\Phi}_c]}
{\delta {\Phi}_c(x)}
\right) = 0, \quad y \gsim x;
\l{3.12c}
\eeq

4. Yang-Feldman relation.
\beq
\int dy
\frac{\delta^2 I}{\delta {\Phi}_c(x) \delta {\Phi}_c(y)}
[\underline{\Phi}_R(y|J) - {\Phi}_c(y|J)] =
ih \underline{W}^+[{\Phi}_c]
\frac{\delta
\underline{W}[{\Phi}_c]}{\delta {\Phi}_c(x)}.
\l{3.12d}
\eeq

5. Boundary condition.
\beq
\underline{W}^+[{\Phi}_c] \hat{\varphi}_h(x) \sqrt{h}
\underline{W}[{\Phi}_c] =
\underline{\Phi}_R(x|J_{\Phi_c}), \quad
x \gsim supp {\Phi}_c,
\l{3.12e}
\eeq
Here $\hat{\varphi}_h(x)  =  \underline{\Phi}_R(x|0)$
is the field operator without source.

\subsection{Set of axioms of covariant semiclassical field theory}

The covariant axioms of semiclassical field theory are as follows.

{\bf C1.} {\it A Hilbert state space $\cal F$ is given.}

{\bf C2.} {\it
An unitary represatation of the Poincare group is given. The operators
of the representation
$\underline{U}_g: {\cal  F}  \to  {\cal  F}$ are asymptoitc
series in $\sqrt{h}$.
}

{\bf C3.} {\it
To each classical source
$J(x)$   with compact support one assignes a retarded field
(LSZ R-function). It is an operator-valued distribution
$\underline{\Phi}_R(x|J)$ expanded in
$\sqrt{h}$ according to
\r{3.1aa}.  It satisfies the properties
\r{3.3o}, \r{3.3a}, \r{3.3b}, \r{3.3c}.
}

{\bf C4.} {\it
To each classical field configuration
$\Phi_c(x)$ with  compact  support  one  assigns  a c-number.  It is a
classical action
$I[\Phi_c]$ satisfying the locality condition
\r{3.11a}.  The property
$\Phi_c(x) = \Phi_R(x|J)$ is valid iff
\beq
J(x) = - \frac{\delta I[\Phi_c]}{\delta \Phi_c(x)}.
\l{3.11f}
\eeq
}

{\bf C5.} {\it
To each classical field configuration $\Phi_c(x)$ with compact support
one  assigns  the   operator   $\underline{W}[\Phi_c]$   expanded   in
$\sqrt{h}$.   It   satisfies   the   relations  \r{3.12a},  \r{3.12b},
\r{3.12c}, \r{3.12d}, \r{3.12e}.
}

\section{Leading order of semiclassical expansion: scalar field}

\subsection{Leading order for semiclassical axioms}

Consider the simplest scalar field model with action
\bez
I[\Phi_c] = \int dx \left[
\frac{1}{2} \partial_{\mu} \Phi_c \partial_{\mu} \Phi_c - V(\Phi_c)
\right],
\eez
satisfying the locality  condition  \r{3.11a}.  For  this  model,  the
classical retarded field is a solution of the Cauchy problem \r{3.2a}.
Denote by $U_g$ the operator $\underline{U}_g$ in the leading order of
perturbation  theory,  while  $\Phi_R^{(1)}(x|J)$  will  be  the first
correction. Denota also $\hat{\varphi}_0(x) \equiv \Phi_R^{(1)}(x|0)$,
let  $W_0[\Phi_c]$ be a scattering operator in the external background
$\underline{W}[\Phi_c]$ in the leading order in $\sqrt{h}$:
\bey
\underline{U}_g \simeq U_g,
\qquad
\underline{W}\left[\Phi_c\right] \simeq {W}_0\left[\Phi_c\right],
\\
\underline{\Phi}_R(x|J) \simeq
\overline{\Phi}_R(x|J) + \sqrt{h} \Phi_R^{(1)}(x|J),
\qquad
\hat{\varphi}_0(x) = \Phi^{(1)}_R(x|0).
\eey
Write down postulate C2 and properties \r{3.3o}, \r{3.3a}, \r{3.3b} of
postulate C3 in the leading order in $\sqrt{h}$:
\beb
U_{g_1g_2} = U_{g_1}U_{g_2},\quad
U_{g^{-1}} \Phi_R^{(1)} (x|u_gJ) U_g = \Phi_R^{(1)}(w_gx|J);
\\
(\Phi_R^{(1)}(x|J))^+ = \Phi_R^{(1)}(x|J),
\quad
\frac{\delta \Phi_R^{(1)}(x|J)}{\delta J(y)} = 0, \quad y \gsim x.
\l{4.1}
\eeb
Commutation relation \r{3.3c} has the following form  in  the  leading
order in $\sqrt{h}$:
\beq
\left[ \Phi_R^{(1)}(x|J), \Phi_R^{(1)}(y|J) \right]
= - i \left(
\frac{\delta \Phi_R(x|J)}{\delta J(y)}
- \frac{\delta \Phi_R(y|J)}{\delta J(x)}
\right)
\l{4.1a}
\eeq
The right-hand side of relation
\r{4.1}  is presented via the retarded Green function
$D^{ret}(x,y|J)$ of equation
\beq
(\partial_{\mu} \partial^{\mu} + V'{}'(\Phi_R(x|J))) \delta \Phi(x)
= \delta J(x),
\quad \delta \Phi|_{x<supp \delta J} = 0
\l{4.2}
\eeq
as follows:
\beq
\left[ \Phi_R^{(1)}(x|J), \Phi_R^{(1)}(y|J) \right]
= - i \left(
D^{ret}(x,y|J) - D^{ret}(y,x|J)
\right)
\l{4.1b}
\eeq
Postulate C5 has the following form in the leading order in $\sqrt{h}$:
\beb
U_g W_0[\Phi_c] U_{g^{-1}} = W_0[u_g\Phi_c],
\quad
W_0^+[\Phi_c] = (W_0[\Phi_c])^{-1},
\\
\frac{\delta}{\delta \Phi_c(y)}
\left(
W_0^+[\Phi_c] \frac{\delta W_0[\Phi_c]}{\delta \Phi_c(x)}
\right) = 0, y\gsim x;
\l{4.3}
\eeb
\beq
(\partial_{\mu} \partial^{\mu} + V'{}'(\Phi_R(x|J)))
\Phi_R^{(1)}(x|J_{\Phi_c}) = 0;
\l{4.3a}
\eeq
\beq
\Phi_R^{(1)}(x|J_{\Phi_c}) =
W_0^+[\Phi_c] \hat{\varphi}_0(x)
W_0[\Phi_c],
\quad x \gsim supp \Phi_c.
\l{4.3b}
\eeq
If the  objects $U_g$,  $\Phi_R^{(1)}(x|J)$,  $W_0[\Phi_c]$ satisfying
eqs.  \r{4.1},  \r{4.1b}, \r{4.3}, \r{4.3a}, \r{4.3b} are specified we
will  say  that  the  leading  order  of semiclassical field theory is
constructed.

\subsection{Free case}

Consider the case of free field theory:
$V(\Phi_c)=  \frac{m^2\Phi_c^2}{2}$.  According to eq.
\r{3.2a}, the semiclassical retarded field is
\bez
\Phi_R(x|J) = \int D_0^{ret}(x,y) J(y) dy,
\eez
where $D_0^{ret}(x,y) = D_0^{ret}(x-y)$ is a retarded
Green function for the Klein-Gordon equation.
The operator
$\Phi_R^{(1)}(x|J)$ satisfies equation \r{4.3a}
\beq
(\partial_{\mu} \partial^{\mu} + m^2) \Phi_R^{(1)}(x|J) = 0.
\l{4.3d}
\eeq
It follows from the Bogoliubov causality condition that
$\Phi_R^{(1)}(x|J)$ does not depend on
$J(y)$  as  $y\gsim  x$.
Since the solution of Klein-Gordon equation
\r{4.3d}  is uniquely expressed via the
initial conditions at $-\infty$, the causality property implies that
$\Phi_R^{(1)}(x|J)$ does not depend on $J$:
\bez
\Phi_R^{(1)}(x|J) = \hat{\varphi}_0(x).
\eez
The field
$\hat{\varphi}_0(x)$ can be identified with quantum free field.
It satisfies the Klein-Gordon equation and commutation relation
\r{4.1b}:
\beb
(\partial_{\mu} \partial^{\mu} + m^2) \hat{\varphi}_0(x) = 0,
\quad \hat{\varphi}_0^+(x) = \hat{\varphi}_0(x),
\\
\left[\hat{\varphi}_0(x), \hat{\varphi}_0(y)\right] =
- i (D_0^{ret}(x,y) - D_0^{ret}(y,x)).
\l{4.3e}
\eeb
It is remarkable that the commutation relations of free  field  theory
are obtained axiomatically,  without using any postulates of canonical
quantization.

It is well-known that  it  is  possible  to  construct  non-equivalent
representations of the canonical commutation relation
\c{3,21}.
However, if  we  suppose  in addition that there exists a vacuum state
vector invariant under Poincare transformations, the representation of
commutation relations
\r{4.3e} will be fixed.
$\hat{\varphi}_0$ should be a free scalar field of the mass
$m$. It is presented via creation and annihilation operators
$a^{\pm}_{{\bf p}}$   of scalar particles with
momentum ${\bf  p}$  as
\c{1,22}
\beq
\hat{\varphi}_0(x) = \frac{1}{(2\pi)^{d/2}}
\int \frac{d{\bf p}}{\sqrt{2\omega_{\bf p}}}
[a_{\bf p}^+ e^{-i{\bf p}{\bf x} + i \omega_{\bf p}t}
+ a_{\bf p}^- e^{i{\bf p}{\bf x} - i \omega_{\bf p}t}].
\l{4.4b}
\eeq
Here $d$ is a number of space dimensions,
$\omega_{\bf  p}  =  \sqrt{{\bf p}^2 + m^2}$.

The properties of the Poincare transformation operator
$U_g$ are taken to the form
\beq
U_{g_1g_2} = U_{g_1} U_{g_2},
\qquad
U_{g^{-1}} \hat{\varphi}_0(x) U_g = \hat{\varphi}_0(w_gx).
\l{4.4c}
\eeq
Therefore, the operator
$U_g$   coincides with Poincare transformation in the
free field theory.

Finally, property
\r{4.3b} for the scattering operator is taken to the form:
\bez
\hat{\varphi}_0(x) W_0[\Phi_c] = W_0[\Phi_c] \hat{\varphi}_0(x).
\eez
This means     that    the    operator    $W_0$    is    a    c-number
$e^{i\gamma[\Phi_c]}$.  The real functional $\gamma[\Phi_c]$ should be
Poincare    invariant    and    satisfy    the    locality    property
$\frac{\delta^2\gamma}{\delta \Phi_c(x) \delta \Phi_c(y)} =  0$.  This
arbitrariness  is  related  with  the possibility of adding a one-loop
correction to the free theory Lagrangian.

Since the renormalization is not necessary, one can set
$\gamma = 0$. Then
\bez
W_0[\Phi_c] = 1.
\eez
For the   free  case,  the  pertubation  series  for  the  objects  of
semiclassical theory contains the finite number of terms:
\bey
\underline{\Phi}_R(x|J) =  \int  D_0^{ret}(x,y)  J(y)  dy  +  \sqrt{h}
\hat{\varphi}_0(x); \\
\underline{U}_g = U_g, \qquad
\underline{W} [\Phi_c] = 1
\eey

\subsection{Retarded semiclassical field of the interaction theory}

Let us investigate the general case, the theory with the potential
$V(\Phi_c)$. As $J=0$ and $\Phi_c=0$, the properties
\r{4.1}, \r{4.1b}, \r{4.3}, \r{4.3a} and \r{4.3b}
coincided with the free case for the square mass
$m^2 = V'{}'(0)$. Therefore,
\bez
\Phi_R^{(1)} (x|0) = \hat{\varphi}_0(x), \quad
W[0]=1,
\eez
while the leading order of Poincare transformations
$U_g$   coincide with free case.

Let $\Phi_c\ne 0$. Denote
\bez
v(x) \equiv V'{}'(\Phi_c(x)) - m^2,
\quad
\hat{\varphi}_v(x)
\equiv \Phi_R^{(1)}(x|J_{\Phi_c}).
\eez
Then eq.\r{4.3a} and boundary condition for the field
$\hat{\varphi}_v(x)$ will be written as
\beq
(\partial_{\mu} \partial^{\mu} + m^2 + v(x))
\hat{\varphi}_v(x) = 0,
\quad
\hat{\varphi}_v(x)|_{x\lsim supp v} = \hat{\varphi}_0(x).
\l{4.5}
\eeq
Relation \r{4.5}  specifies the field $\Phi_R^{(1)}(x|J)$ as $J\sim 0$
uniquely. Making use of the Bogoliubov causality property, one extends
this definition to the arbitrary case. Properties \r{4.1} of the field
$\Phi_R^{(1)}(x|J)$ are corrolaries of the construction.

Commutation relation
\r{4.1b} to be checked can be rewritten as
\beq
[\hat{\varphi}_v(x), \hat{\varphi}_v(y)] = \frac{1}{i} D_v(x,y),
\qquad D_v(x,y) = D_v^{ret}(x,y) - D_v^{adv}(x,y),
\l{4x.7}
\eeq
where $D_v^{ret}(x,y)$ and
$D_v^{adv}(x,y) \equiv D_v^{ret}(y,x)$ are
retarded and advanced Green functions for eq.\r{4.5}.
They satisfy the relation
\bez
(\partial_{\mu}^x \partial^{\mu}_x + m^2 + v(x))
D_v^{ret,adv} (x,y) = \delta(x-y),
\eez
moreover, the function $D_v^{ret}(x,y)$ vanishes as $x\lsim y$,  while
$D_v^{adv}(x,y)$ vanishes as $x\gsim y$.

To check commutation relation \r{4x.7}, consider the difference
\bez
C_v(x,y) =
[\hat{\varphi}_v(x), \hat{\varphi}_v(y)] - \frac{1}{i} D_v(x,y).
\eez
It satisfies eq.
\r{4.5} with respect to $x$ and $y$:
\bez
[\partial_{\mu}^x \partial^{\mu}_x + m^2 + v(x)]
C_v(x,y) = 0,
\quad
[\partial_{\mu}^y \partial^{\mu}_y + m^2 + v(y)]
C_v(x,y) = 0.
\eez
It vanishes as $x,y < supp v$,  since in this domain $\hat{\varphi}_v$
coincides   with   the   free   field.  $\hat{\varphi}_0$.  Therefore,
$C_v(x,y)=0$, and property \r{4x.7} is checked.

\subsection{Asymptotic condition, Relations for scattering operator}

Since $v(x)$ is a function  with  compact  support,  for  the  domains
$x\gsim  supp  v$  and  $x\lsim supp v$ eq.\r{4.5} coincides with free
field equation. Therefore, fields $\hat{\varphi}_v(x)$ as $x\gsim supp
v$  and  as  $x\lsim supp v$ coincide with asymptotic out- and in-free
fields $\hat{\varphi}^v_{out}(x)$  and  $\hat{\varphi}_{in}(x)  \equiv
\hat{\varphi}_0(x)$. Asymptotic fields satisfy Klein-Gordon equations.

To represent field
$\hat{\varphi}_v(x)$ via the asymptotic in-field,
notice that the difference satisfy the following equation:
\bez
(\partial_{\mu} \partial^{\mu} + m^2 + v(x))
(\hat{\varphi}_{v}(x) - \hat{\varphi}_{in}(x)) =
- v(x) \hat{\varphi}_{in}(x))
\eez
It vanishes as $x\lsim supp v$; therefore,
\beq
(\hat{\varphi}_{v}(x) - \hat{\varphi}_{in}(x)) =
- \int dy D^{ret}_v(x,y) v(y)
\hat{\varphi}_{in}(y)).
\l{4x.2}
\eeq
This formula can be presentes  in  another  form.  Consider  one  more
equation for the difference:
\bez
(\partial_{\mu} \partial^{\mu} + m^2)
(\hat{\varphi}_{v}(x) - \hat{\varphi}_{in}(x)) =
- v(x) \hat{\varphi}_{v}(x)).
\eez
It implies that
\beq
(\hat{\varphi}_{v}(x) - \hat{\varphi}_{in}(x)) =
- \int dy D^{ret}_0(x,y) v(y)
\hat{\varphi}_{v}(y)).
\l{4x.3}
\eeq
It is convenient to write formulas \r{4x.2} and \r{4x.3} in a symbolic
form:
\beq
\hat{\varphi}_v = (1-D_v^{ret}v)\hat{\varphi}_{in},
\quad
\hat{\varphi}_{in} = (1+D_0^{ret}v)\hat{\varphi}_{v}.
\l{4x.4}
\eeq
Here $D_v^{ret}$ is the operator with kernel $D_v^{ret}(x,y)$,  $v$ is
the  operator  of  multiplication by $v(x)$.  It follows from \r{4x.4}
that the operators $(1-D_v^{ret}v)$ and $(1+D_0^{ret}v)$ are  inverse,
i.e.
\beq
D_v^{ret} = (1+D_0^{ret}v)^{-1} D_0^{ret}.
\l{4x.5}
\eeq
Analogs of the formulas
\r{4x.4} for asymptotic out-fields are
\beq
\hat{\varphi}_v = (1-D_v^{adv}v)\hat{\varphi}_{out},
\quad
\hat{\varphi}_{out} = (1+D_0^{adv}v)\hat{\varphi}_{v}.
\l{4x.6}
\eeq

Property \r{4.3b} can be written for the unitary operator $W_0[\Phi_c]
\equiv W_0\{v\}$:
\beq
\hat{\varphi}^v_{out} =
W_0^+\{v\} \hat{\varphi}_{in} W_0\{v\}.
\l{4y.1}
\eeq
We see that the operator $W_0\{v\}$ is an S-matrix for  quantum  field
theory  in the external background $v(x)$.  Let us take eq.\r{4y.1} to
the more general form.

Free fields $\hat{\varphi}^v_{out}(x)$ and $\hat{\varphi}_{in}(x)$ are
expanded to positive- and negative-frequency parts:
\beb
\hat{\varphi}_{in,out} (x) =
\hat{\varphi}_{in,out}^+ (x) +
\hat{\varphi}_{in,out}^- (x),
\\
\hat{\varphi}_{in}^{\pm} (x) =
\frac{1}{(2\pi)^{d/2}} \int
\frac{d{\bf p}}{\sqrt{2\omega_{\bf p}}}
a_{\bf p}^{\pm} e^{\pm i (\omega_{\bf p} t - {\bf p}{\bf x})},
\\
\hat{\varphi}_{out}^{\pm} (x) =
\frac{1}{(2\pi)^{d/2}} \int
\frac{d{\bf p}}{\sqrt{2\omega_{\bf p}}}
b_{\bf p}^{\pm} e^{\pm i (\omega_{\bf p} t - {\bf p}{\bf x})}.
\l{4y.2}
\eeb
It follows   from   eqs.\r{4x.4}   and   \r{4x.6}   that   the   field
$\hat{\varphi}_{out}(x)$     is     a     linear     combination    of
$\hat{\varphi}_{in}(x)$.  Therefore,  the dependence of  creation  and
annihilation  operators for out-particles $b_{\bf p}^{\pm}$ of $a_{\bf
p}^{\pm}$ is also linear:
\beb
b_{\bf p}^{+} = a_{\bf p}^{+} + \int d{\bf k}
(A_{\bf  pk} a^+_{\bf k} + B_{\bf pk}a^-_{\bf k}),
\\
b_{\bf p}^{-} = a_{\bf p}^{-} + \int d{\bf k}
(A_{\bf  pk}^* a^-_{\bf k} + B_{\bf pk}^* a^+_{\bf k}),
\l{4y.4}
\eeb
or in symbolic form
\beb
b^+ = a^+ + Aa^+ + Ba^-,
\qquad
b^- = a^- + A^*a^- + B^*a^+,
\l{4y.5}
\eeb
where $A,B,A^*,B^*$ are operators with kernels $A_{\bf  pk}$,  $B_{\bf
pk}$, $A^*_{\bf pk}$, $B^*_{\bf pk}$ correspondingly.

Relation \r{4y.1}  can  be  viewed  as  a  transformation of operators
$a_p^{\pm}$ and $b_p^{\pm}$ that conserves the  canonical  commutation
relations.  According  to  the  Berezin  theorem  \c{22},  a canonical
transformation is unitary iff the  function  $B_{\bf  pk}$  is  square
integrable.

Find an  explicit  form of the function $B_{\bf pk}$ and show that the
Berezin condition is satisfied.  It follows from \r{4x.4} and \r{4x.6}
that
\bez
\hat{\varphi}_{out} - \hat{\varphi}_{in} =
[(1+D_0^{adv}v) (1-D_v^{ret}v) - 1] \hat{\varphi}_{in} =
- (1+D_0^{adv}v)D_vv \hat{\varphi}_{in}.
\eez
In particular, for the case $x\gsim supp v$ we write:
\bez
\hat{\varphi}_{out}(x) = \hat{\varphi}_{in}(x)
- \int D_v(x,y) v(y) \hat{\varphi}_{in}(y) dy.
\eez
The operator $b^+_{\bf p}$ can be expressed via  $\hat{\varphi}_{out}$
and present as an integral over surface $x^0=сonst$ as $x>supp v$:
\bey
b^+_{\bf p}  =
\frac{1}{(2\pi)^{d/2}} \sqrt{\frac{\omega_{\bf p}}{2}}
\int_{x^0=const} d{\bf x}
e^{-i(\omega_{\bf p}x^0 - {\bf p}{\bf x})}
\left\{
\hat{\varphi}_{out}(x) -
\frac{i}{\omega_{\bf p}}
\frac{\partial}{\partial x^0}
\hat{\varphi}_{out}(x)
\right\} =
\\
a_p^+ -
\frac{1}{(2\pi)^{d/2}} \sqrt{\frac{\omega_{\bf p}}{2}}
\int_{x^0=const} d{\bf x} \int dy
e^{-i(\omega_{\bf p}x^0 - {\bf p}{\bf x})}
\left\{
1 - \frac{i}{\omega_{\bf p} \frac{\partial}{\partial x^0}}
\right\}
D_v(x,y) v(y)
\hat{\varphi}_{in}(y).
\eey
Therefore, we find $B_{\bf pk}$:
\bey
B_{\bf pk} =
- \frac{1}{(2\pi)^d} \frac{1}{2}
\sqrt{\frac{\omega_{\bf p}}{\omega_{\bf k}}}
\int_{x^0=const} d{\bf x} \int dy
e^{-i
(\omega_{\bf p}x^0 - {\bf p}{\bf x})
- i (\omega_{\bf k}y^0 - {\bf k}{\bf y})
}
\\ \times
\left\{
1 - \frac{i}{\omega_{\bf p} \frac{\partial}{\partial x^0}}
\right\}
D_v(x,y) v(y)
\eey
This function rapidly damps at infinity  as  a  Fourier  transform  of
a smooth   function  with  compact  support.  Therefore,  the  Berezin
condition is satisfied, so that an unitary operator
$W_0\{v\}$ is defined from \r{4y.1} up to a c-number multiplier.

To check  the  properties  \r{4.3}  and  construct  the  semiclassical
theory, one should develop the calculus of normal symbols.

\subsection{Calculus of normal symbols. Propagator}

A calculus of normal symbols is often used in quantum field theory
\c{5,7}. The  physical  quantities are presented as a series in normal
products of the fields. They multiplies according to the Wick theorem.

If there is an external background, the normal products can be defined
in  different ways.  The simplest way is the following:  one expresses
the field  $\hat{\varphi}_v$  via  the  free  in-field  at  $-\infty$,
expands  it  into  parts with creation and with annihilation operators
$a^{\pm}$ and puts the creation operators to the left and annihilation
operators to the right.

However, in  order to make the semiclassical perturbation theory to be
analogous to usual Feynman theory,  it is more convenient  to  express
field  $\hat{\varphi}_v$  via negative-frequency component of in-field
$\hat{\varphi}_{in}^-$ and positive-frequency component  of  out-field
$\hat{\varphi}_{out}^+$; then one puts the operators $b^+$ to the left
and $a^-$ to the right.

Under this definition, the average value of normal product of operators
with respect  to  in  and  out vacuums will be nonzero.  However,  the
matrix element between vacuums
\beq
<A> \equiv
\frac{<0,out|A|0,in>}{<0,out|0,in>} =
\frac{<0|W_0\{v\}A|0>}{<0|W_0\{v\}|0>},
\l{4z.3}
\eeq
will vanish:
\bez
<:\hat{\varphi}_v(x_1)...\hat{\varphi}_v(x_k):> = 0.
\eez

\subsubsection{Expansion of the field}

Express the operator
$\hat{\varphi}_v$  via
$\hat{\varphi}_{in}^-$ and $\hat{\varphi}_{out}^+$.
If there are no external fields, one has
$\hat{\varphi}_v = \hat{\varphi}_{in}^- + \hat{\varphi}_{out}^+$.
Therefore, one should look the expansion in the form
\bez
\hat{\varphi}_v = \hat{\varphi}_{in}^- + \hat{\varphi}_{out}^+ +
\Delta \hat{\varphi}.
\eez
The field $\Delta \hat{\varphi}$ satisfies the follwing equation:
\bez
(\partial_{\mu}\partial^{\mu} + m^2 + v(x)) \Delta \hat{\varphi}(x) =
- v(x)
(\hat{\varphi}_{in}^-(x) + \hat{\varphi}_{out}^+(x))
\eez
and boundary condition at $-\infty$ and $+\infty$:
\bey
\Delta \hat{\varphi}(x) =
\hat{\varphi}_{in}^+(x) -
\hat{\varphi}_{out}^+(x), \quad
x \lsim supp v,
\\
\Delta \hat{\varphi}(x) =
\hat{\varphi}_{out}^-(x) -
\hat{\varphi}_{in}^-(x), \quad
x \gsim supp v.
\eey
Therefore, field    $\Delta    \hat{\varphi}(x)$     contains     only
negative-frequency   part   at  $+\infty$  and  positive-frequency  at
$-\infty$. These boundary conditions are called Feynman. By $D^c(x,y)$
we   denote   the  causal  Green  function  of  equation  \r{4.5}  (or
propagator). It satisfies the relation
\bez
(\partial_{\mu}^x \partial^{\mu}_x   +   m^2   +  v(x))  D_v^c(x,y)  =
\delta(x-y)
\eez
and Feynman boundary conditions. Therefore,
\bez
\Delta \hat{\varphi}(x) = - \int dy D_v^c(x,y) v(y)
(\hat{\varphi}_{in}^-(y) + \hat{\varphi}_{out}^+(y)),
\eez
or in symbolic form
\beq
\hat{\varphi}_v =
(1-D_v^cv)(\hat{\varphi}_{in}^-
+ \hat{\varphi}_{out}^+).
\l{4x.7a}
\eeq
Analogously to formula \r{4x.2},  one can write eq.\r{4x.7a} as
\beq
\hat{\varphi}_v =
(1-D_0^cv)^{-1}(\hat{\varphi}_{in}^-
+ \hat{\varphi}_{out}^+).
\l{4x.7b}
\eeq
Therefore,
\beq
D_v^c = D_0^c (1+vD_0^c)^{-1}.
\l{4x.7c}
\eeq
Let us  show  that  there  exists  a Green function that satisfies the
Feynman boundary conditions.

\subsubsection{Existence of propagator}

Consider the problem  of  finding  the  function  $f$  satisfying  the
Feynman boundary condition and equation
\beq
(\partial_{\mu}\partial^{\mu} + m^2 + v(x)) f(x) = g(x),
\l{4y.6a}
\eeq
with fixed right-hand side $g(x)$ with compact support.  Consider  the
difference $\tilde{f} = f - D_v^{ret}g$ satisfying equation
\bez
(\partial_{\mu}\partial^{\mu} + m^2 + v(x)) \tilde{f}(x) = 0.
\eez
Therefore, the function
$\tilde{f}$ can be expressed as
\bey
\tilde{f}(x) =
\frac{1}{(2\pi)^{d/2}} \int \frac{d{\bf p}}{\sqrt{2\omega_{\bf p}}}
\left[
\overline{\beta}_{\bf p} e^{i(\omega_{\bf p}x^0 - {\bf p}{\bf x})}
+
{\beta}_{\bf p} e^{i(\omega_{\bf p}x^0 - {\bf p}{\bf x})}
\right],
\quad x \gsim supp v,
\\
\tilde{f}(x) =
\frac{1}{(2\pi)^{d/2}} \int \frac{d{\bf p}}{\sqrt{2\omega_{\bf p}}}
\left[
\overline{\alpha}_{\bf p} e^{i(\omega_{\bf p}x^0 - {\bf p}{\bf x})}
+
{\alpha}_{\bf p} e^{i(\omega_{\bf p}x^0 - {\bf p}{\bf x})}
\right],
\quad x \lsim supp v,
\eey
with coefficient  functions   $\alpha_{\bf   p}$,   $\beta_{\bf   p}$,
$\overline{\alpha}_{\bf p}$, $\overline{\beta}_{\bf p}$.
Analogously to derivation of formula
\r{4y.5}, we find the following relations:
\beq
\overline{\beta} = \overline{\alpha} + A\overline{\alpha} + B\alpha,
\qquad
{\beta} = {\alpha} + A^*{\alpha} + B^*\overline{\alpha}.
\l{4y.6}
\eeq
The function $D^{ret}g$ is given.  Write it expansion  into  frequency
components as $x \gsim supp v$ and $x \gsim supp g$:
\bez
\tilde{f}(x) =
\frac{1}{(2\pi)^{d/2}} \int \frac{d{\bf p}}{\sqrt{2\omega_{\bf p}}}
\left[
{\gamma}^*_{\bf p} e^{i(\omega_{\bf p}x^0 - {\bf p}{\bf x})}
+
{\gamma}_{\bf p} e^{i(\omega_{\bf p}x^0 - {\bf p}{\bf x})}
\right].
\eez
The boundary conditions for $f$ will be satisfied iff
\bez
\beta_{\bf p} + \gamma_{\bf p} = 0,
\quad \overline{\alpha}_{\bf p} = 0.
\eez
It follows from  \r{4y.6}  that  $\alpha  =  -  (1+A^*)^{-1}  \gamma$,
therefore,  the  solution  of the problem \r{4y.6a} and Green function
for this problem are uniquely specified. The operator
$(1+A^*)$ is  invertible  because  of  general  properties  of  linear
canonical transformations \c{22}.

\subsubsection{Normal form of the product of operators}

Analogously to the free case, the product of operators
$\hat{\varphi}_v(x) \hat{\varphi}_v(y)$
can be presented in the following normal form:
\beq
\hat{\varphi}_v(x) \hat{\varphi}_v(y)
=
:\hat{\varphi}_v(x) \hat{\varphi}_v(y): +
<\hat{\varphi}_v(x) \hat{\varphi}_v(y)>,
\l{4z.1}
\eeq
hear the "correlator" of the fields
$<\hat{\varphi}_v(x) \hat{\varphi}_v(y)>$
is a c-number function
\beq
<\hat{\varphi}_v(x) \hat{\varphi}_v(y)> =
[(1+D_0^cv)^{-1}\hat{\varphi}_{in}^-(x),
(1+D_0^cv)^{-1}\hat{\varphi}_{out}^+(y)].
\l{4z.2}
\eeq

\subsubsection{
Normal form of T-product
}

It is possible to obtain an analogous result for the T-product:
\bey
T \hat{\varphi}_v(x) \hat{\varphi}_v(y) \equiv
\hat{\varphi}_v(x) \hat{\varphi}_v(y)  -
\theta(y^0-x^0)
[\hat{\varphi}_v(x), \hat{\varphi}_v(y)] =
\\
:\hat{\varphi}_v(x) \hat{\varphi}_v(y): +
\frac{1}{i} D_v^c(x,y).
\eey

\subsubsection{Hermite conjugation}

It is important to note that the normal product
$:\hat{\varphi}_v(x) \hat{\varphi}_v(y):$
is not a Hermitian operator. Namely, find a conjugated operator
$(:\hat{\varphi}_v(x) \hat{\varphi}_v(y):)^+$.
Write the relation
\r{4z.1} and its conjugation:
\bey
\hat{\varphi}_v(x) \hat{\varphi}_v(y) =
:\hat{\varphi}_v(x) \hat{\varphi}_v(y): +
\frac{1}{i} (D_v^c - D_v^{adv})(x,y),
\\
\hat{\varphi}_v(y) \hat{\varphi}_v(x) =
(:\hat{\varphi}_v(x) \hat{\varphi}_v(y):)^+ -
\frac{1}{i} (D_v^{c*} - D_v^{adv})(x,y).
\eey
Therefore,
\bez
\frac{1}{i} D_v(x,y) =
:\hat{\varphi}_v(x) \hat{\varphi}_v(y): -
(:\hat{\varphi}_v(x) \hat{\varphi}_v(y):)^+
+ \frac{1}{i} (D_v^c + D_v^{c*} - 2D_v^{adv})(x,y).
\eez
Thus,
\beb
(:\hat{\varphi}_v(x) \hat{\varphi}_v(y):)^+ =
:\hat{\varphi}_v(x) \hat{\varphi}_v(y):
+ \frac{1}{i} \Delta_v(x,y),
\\
\Delta_v = D_v^c + D_v^{c*} - D_v^{ret} - D_v^{adv}.
\l{4x.11}
\eeb

\subsubsection{Variation of matrix elements}

Obtain an expression for the derivative of the matrix element
\r{4z.3}. Let us variate the relation
\bez
<0|W_0\{v\}A|0> = <0|W_0\{v\}|0> <A>
\eez
with respect to $v(x)$. We obtain:
\beb
<0|\frac{\delta W_0}{\delta v(x)} A |0>
+
<0|W_0 \frac{\delta A}{\delta v(x)}|0>
= \\
<0|\frac{\delta W_0}{\delta v(x)}|0> <A>
+ <0|W_0|0> \frac{\delta}{\delta v(x)} <A>.
\l{4z.4}
\eeb
It is convenient to introduce the following notations:
\beq
\hat{\rho}(x|v) = \frac{1}{i} W_0^+\{v\}
\frac{\delta W_0\{v\}}{\delta v(x)}.
\l{4z.5}
\eeq
Within the Bogoliubov S-matrix framework,  this is a current operator.
Under notations \r{4z.5}, relation \r{4z.4} is taken to the form:
\beq
<\frac{\delta A}{\delta v(x)}>
\frac{\delta}{\delta v(x)} <A>
- i (<\hat{\rho}(x|v) A> - <\hat{\rho}(x|v)><A>).
\l{4z.6}
\eeq

\subsection{Investigation of Poincare covariance, unitarity, causality}

Investigate whether  the  operator  $W_0\{v\}$  specified   from   the
relation
\beq
\hat{\varphi}_{in}(x) W_0\{v\} = W_0\{v\} \hat{\varphi}_{out}^v
\l{4u.0}
\eeq
satisfy the properties of Poincare invariance, unitarity, causality
\r{4.3}.   Denote
\beq
<0|W_0\{v\}|0> = e^{iA_1\{v\}}.
\l{4u.1}
\eeq
The quantity $A_1\{v\}$ is a one-loop correction to  effective  action
of  the  theory.  Making  use  of  $A_1\{v\}$,  one finds the operator
$W_0\{v\}$ from relation \r{4u.0} uniquely.

Properties of Poincare invariance,  unitarity and  causality  requires
additional  conditions  on  the  one-loop effective action $A_1\{v\}$.
Find them.

The property of Poincare invariance
of the  operator  $W_0$  leads to Poincare invariance of the effective
action:
\beq
A_1\{u_gv\} = A_1\{v\}.
\l{4u.1a}
\eeq

Investigate the unitarity property. It follows from
\r{4u.0} that
\bez
[\hat{\varphi}_{out}(x), W_0^+W_0] = 0,
\qquad
[\hat{\varphi}_{in}(x), W_0W_0^+] = 0.
\eez
Therefore, the operators
$W_0W_0^+$   and  $W_0^+W_0$  are proportional to unit operator.
To check the unitarity, it is sufficient to check that
\beq
<0|W_0^+\{v\} W_0\{v\}|0> = 0.
\l{4u.2}
\eeq
It is satisfied as $v=0$. Therefore, it is sufficient to check that
\bez
\frac{\delta}{\delta v(y)}
<0|W_0^+\{v\}W_0\{v\}|0> = 0,
\eez
or in notations \r{4z.5}
\beq
<0|\hat{\rho}(y|v)|0> =
(<0|\hat{\rho}(y|v)|0>)^*.
\l{4u.3}
\eeq
The causality condition can be presented as
\beq
\frac{\delta \hat{\rho}(y|v)}{\delta v(x)} = 0,
\qquad x \gsim y.
\l{4u.4}
\eeq

To present properties \r{4u.3}  and  \r{4u.4}  as  conditions  on  the
one-loop  effective  action  $A_1\{v\}$,  find an explicit form of the
operator  $\hat{\rho}(y|v)$.  Consider  the  variation   of   relation
\r{4u.0} with respect to $v$. Take into account that the field
$\hat{\varphi}_{in}(x)  =  \hat{\varphi}_0(x)$  does not depend on
$v$, while
$\frac{\delta W_0}{\delta v(y)} = i W_0\hat{\rho}(y|v)$. Then
\bez
\hat{\varphi}_0(x) i W_0\{v\} \hat{\rho}(y|v) =
i W_0\{v\} \hat{\rho}(y|v) \hat{\varphi}_{out}(x) +
W_0\{v\}
\frac{\delta \hat{\varphi}_{out}^v}{\delta v(y)},
\eez
or
\beq
\frac{\delta \hat{\varphi}_{out}^v(x)}{\delta v(y)}
= i [\hat{\varphi}_{out}^v(x),\hat{\rho}(y|v)].
\l{4u.5}
\eeq
Let us   obtain   from   \r{4u.5}   formula   for  commutator  between
$\hat{\rho}(y|v)$ and $\hat{\varphi}_v(x)$. Use relations \r{4x.4} and
\r{4x.6}:
\bez
\hat{\varphi}_v =
(1+D_0^{ret}v)^{-1} \hat{\varphi}_{in}
(1+D_0^{adv}v)^{-1} \hat{\varphi}_{out}.
\eez
Variating, one finds:
\bey
\delta \hat{\varphi}_v =
- (1+D_0^{ret}v)^{-1}   D_0^{ret}   \delta    v    (1+D_0^{ret}v)^{-1}
\hat{\varphi}_{in} =
- D_v^{ret} \delta v \hat{\varphi}_v,
\\
\delta \hat{\varphi}_v =
- D_v^{adv} \delta v \hat{\varphi}_v +
(1+D_0^{adv}v)^{-1} \delta \hat{\varphi}_{out}.
\eey
Let us use the property \r{4u.5}:
\beb
\frac{\delta \hat{\varphi}_v(x)}{\delta v(y)} =
- D_v^{ret}(x,y) \hat{\varphi}_v(y),
\\
\frac{\delta \hat{\varphi}_v(x)}{\delta v(y)} =
- D_v^{adv}(x,y) \hat{\varphi}_v(y) +
i [\hat{\varphi}_v(x), \hat{\rho}(y|v)].
\l{4u.5a}
\eeb
Therefore, the commutator $\hat{\rho}(y|v)$        and field
$\hat{\varphi}_v(x)$ linearly depnds
on the field $\hat{\varphi}_v$:
\beq
[\hat{\varphi}_v(x), \hat{\rho}(y|v)] = i D_v(x,y) \hat{\varphi}_v(y).
\l{4u.6}
\eeq
Property
\r{4u.6}
specifies the operator $\hat{\rho}(y|v)$ up to a multiplier.
Under notations \r{4z.3},
\beq
<\hat{\rho}(y|v)> =
\frac{\frac{1}{i}<0|\frac{\delta W_0}{\delta v(y)}|0>}
{<0|W_0\{v\}|0>} =
\frac{\delta A_1\{v\}}{\delta v(y)}.
\l{4u.7}
\eeq
Making use of the normal symbol calculus, we find the operator
$\hat{\rho}(y|v)$ satisfying properties
\r{4u.6} and \r{4u.7}:
\beq
\hat{\rho}(y|v) = - \frac{1}{2} :\hat{\varphi}_v(y): +
\frac{\delta A_1\{v\}}{\delta v(y)}.
\l{4u.8}
\eeq

To check the unitarity \r{4u.3}, consider the difference
\bez
\hat{\rho}^+(y|v) - \hat{\rho}(y|v) =
- \frac{1}{2i} \Delta_v(y,y) +
\frac{\delta}{\delta v(y)}
(A_1^*\{v\} - A_1\{v\}).
\eez
The properties $D_v^{ret}(y,y) = 0$ and  $D_v^{adv}(y,y)  =  0$  imply
that   the  function  $\Delta_v(y,y)$  \r{4x.11}  coincide  with  $2Re
D_v^c(y,y)$. Therefore, unitarity is satisfied if
\beq
\frac{\delta}{\delta v(y)} Im A_1\{v\} = \frac{1}{2} Re D_v^c(y,y).
\l{4u.9}
\eeq

To investigate the causality property
\r{4u.4}, one should
show that the commutator
\beq
[\hat{\varphi}_v(z), \frac{\delta \hat{\rho}(y|v)}{\delta v(x)}] = 0
\l{4u.9a}
\eeq
for $x\gsim  y$
vanishes. Then one obtains conditions on the one-loop  effective  action
such that
\beq
<\frac{\delta \hat{\rho}(y|v)}{\delta v(x)}> = 0.
\l{4u.9b}
\eeq
Then properties
\r{4u.9a}  and \r{4u.9b}  imply the Bogoliubov causality property.

1. It follows from the explicit calculation that
\bey
[\hat{\varphi}_v(z), \frac{\delta \hat{\rho}(y|v)}{\delta v(x)}] =
\frac{\delta}{\delta v(x)}
[\hat{\varphi}_v(z), \hat{\rho}(y|v)]
-
[\frac{\delta}{\delta v(x)}
\hat{\varphi}_v(z), \hat{\rho}(y|v)]
\\
=
\frac{\delta}{\delta v(x)} (iD_v(z,y)\hat{\varphi}_v(y))
- D_v^{ret}(z,x) [\hat{\varphi}_v(x), \hat{\rho}(y|v)] =
\\
i \frac{\delta D_v(z,y)}{\delta v(x)}
\hat{\varphi}_v(y) -
i D^{ret}_v(z,x) D^{ret}_v(x,y) \hat{\varphi}_v(y).
\eey
Property
\bey
\frac{\delta D_v(z,y)}{\delta v(x)} = - D_v^{ret}(z,x) D_v^{ret}(x,y),
\qquad x \gsim y
\eey
implies eq. \r{4u.9a}.

2. Property \r{4u.9b} is taken according to
\r{4z.6} to the form:
\bez
\frac{\delta}{\delta v(x)}
<\hat{\rho}(y|v)> =
i(<\hat{\rho}(x|v) \hat{\rho}(y|v)>
- <\hat{\rho}(x|v)><\hat{\rho}(y|v)>),
\eez
or
\bey
\frac{\delta^2A_1\{v\}}{\delta v(x)\delta v(y)} =
\frac{i}{4}
<:\hat{\varphi}_v^2(x): :\hat{\varphi}_v^2(y):> =
\frac{i}{2} (\frac{1}{i} D_v^-(x,y))^2 =
\frac{1}{2i} (D_v^c(x,y))^2, \\ x \gsim y.
\eey
Since the propagator
$D_v^c$ is  symmetric  under  transpositions  of  the  arguments,  the
property
своих аргументов, свойство
\beq
\frac{\delta^2A_1\{v\}}{\delta v(x)\delta v(y)} =
\frac{1}{2i} (D_v^c(x,y))^2, \qquad x \ne y.
\l{4u.11}
\eeq
is satisfied as $x\gsim y$ and $y\gsim  x$;  the  case  $x=y$  is  not
involved to formula \r{4u.11}.

Thus, all  the axioms of the semiclassical field theory are checked in
the leading order  of  perturbation  theory.  It  is  shown  that  the
one-loop effective action allows is to reconstruct the operator
$W_0\{v\}$, relations
\r{4u.1a}, \r{4u.9}   and  \r{4u.11}  for  the  effective  action  are
obtained. These relations shows that the effective  action  cannot  be
chosen to be arbitrary. A one-loop contribution $A_1$ is defined up to
a local functional.

Note that formally
\beq
A_1\{v\} = \frac{i}{2} ln det (1+D_0^cv);
\l{4u.12}
\eeq
then
\beq
\frac{\delta A_1\{v\}}{\delta v(y)} = \frac{i}{2} D_v^c(y,y),
\l{4u.13}
\eeq
and all realtions are satisfied. Formula
\r{4u.12} is in agreement with resummation of Feynman graphs method.
Namely, the S-matrix in the external scalar field
$v(x)$  can be presented as
\bez
W_0\{v\} = Te^{-\frac{i}{2}\int dx v(x) \hat{\varphi}_0^2(x)},
\eez
and vacuum average of this operator is expressed via determinant
\r{4u.12}.

However, definition \r{4u.12} is formal due to divergences.  Relations
\r{4u.1a}, \r{4u.9} and \r{4u.11} may be viewed as a definition of the
renormalized determinant of operator.

Formula \r{4u.13}  may  be  viewed  as  an  additional  condition  for
propagator at coinciding arguments:
\beq
(D_v^c(y,y))_R \equiv \frac{2}{i} \frac{\delta A_1\{v\}}{\delta v(y)}.
\l{4u.14}
\eeq
Relation
\r{4u.11} is used for definition
of square of distribution
$D_v^c(x,y)$ at $x=y$:
\bez
(D_v^c(x,y))^2_R =  2i  \frac{\delta^2  A_1\{v\}}{\delta  v(x)  \delta
v(y)}.
\eez

\subsection{Generalization to multicomponent fields}

Consider the multicomponent fields
$\Phi_c = (\Phi_c^1,...,\Phi_c^k)$.

Let
$I[\Phi_c]$ be classical action,
$I_0[\Phi_c]$ be action of the free theory,
$K^{ij}$ be operator with the kernel
$-\frac{\delta^2I}{\delta \Phi_c^i(x) \delta \Phi_c^j(y)}$,
$K_0^{ij}$ be operator with the kernel
$-\frac{\delta^2I_0}{\delta \Phi_c^i(x) \delta \Phi_c^j(y)}$,
$K' = K - K_0$,
$\hat{\varphi}(x|\Phi_c) \equiv \Phi_R^{(1)}(x|J_{\Phi_c})$.

Eq.\r{4.3a} and boundary condition for the field
$\hat{\varphi}(x|\Phi_c)$ will be written as
\beb
(K_0 + K') \hat{\varphi}(x|\Phi_c) = 0,
\\
\hat{\varphi}(x|\Phi_c) = \hat{\varphi}_0(x), \quad
x\lsim supp \Phi_c,
\l{4m.1}
\eeb
where $\hat{\varphi}_0(x)$ is a free field.

Denote by
$D^{ret|ij}(x,y|\Phi_c)$ the retarded Green function
for eq.\r{4m.1}. It satisfies the relation
\beq
K^{ij} D^{ret|jl} (x,y|\Phi_c) = \delta^{il} \delta(x-y)
\l{4m.2}
\eeq
and the retarded boundary condition
$D^{ret}(x,y|\Phi_c)=0$  as  $x\lsim  y$.  Then
the function
\bez
D^{adv|ij}(x,y|\Phi_c) = D^{ret|ji}(y,x|\Phi_c)
\eez
will satisfy eq.\r{4m.2} with advanced boundary condition, so that
$D^{adv}(x,y|\Phi_c)$    will call as an advanced Green function.

The commutation relations of the fields are written as
\beq
[\hat{\varphi}^i(x|\Phi_c),\hat{\varphi}^j(y|\Phi_c)] =
\frac{1}{i} D^{ij}(x,y|\Phi_c),
\qquad
D^{ij} = D^{ret|ij} - D^{adv|ij}.
\l{4m.3}
\eeq
Analogously to
\r{4x.4},  \r{4x.6}, the solution of eq.
\r{4m.1} can  be  expressed  via  the asymptotic in- and out-fields as
follows:
\beb
\hat{\varphi} = (1-D^{ret}K')\hat{\varphi}_{in} =
(1+ D_0^{ret}K')^{-1} \hat{\varphi}_{in},
\\
\hat{\varphi} = (1-D^{adv}K')\hat{\varphi}_{out} =
(1+ D_0^{adv}K')^{-1} \hat{\varphi}_{out},
\l{4m.4}
\eeb
where $D_0^{ret}$ and $D_0^{adv}$ are advanced and retarded
Green functions of the free field.

The operator $W_0[\Phi_c]$ is specified from the relation:
\beq
\hat{\varphi}_{in}(x) W_0[\Phi_c] = W_0[\Phi_c] \hat{\varphi}_{out}(x)
\l{4m.5}
\eeq
up to a multiplier. It is fixed by a vacuum average:
\beq
<0|W_0[\Phi_c]|0> = e^{iA_1[\Phi_c]}.
\l{4m.5a}
\eeq
Conditions of Poincare invariance,  unitarity and causality  gives  us
restrictions on the one-loop effective action
$A_1[\Phi_c]$.

Formulas of  calculus  of  normal  symbols  can  be   generalized   to
multicomponent case as follows.

Let $D^{c|ij}$  be causal Green function (propagator) of eq.
\r{4m.2}. It contains positive-frequency componants at
$-\infty$  and negative-frequency components
at $+\infty$. Then generalizations of eqs.
\r{4x.7a} and \r{4x.7b} have the form:
\bez
\hat{\varphi} =
(1+D^cK')
(\hat{\varphi}_{in}^- + \hat{\varphi}_{out}^+) =
(1+D^c_0K')^{-1}
(\hat{\varphi}_{in}^- + \hat{\varphi}_{out}^+).
\eez
One also has:
\bez
D^c = D_0^c (1+K'D_0^c)^{-1}.
\eez
Formulas of taking operators to the normal form are:
\bey
<\hat{\varphi}_i(x|\Phi_c) \hat{\varphi}_j(y|\Phi_c)>
= \frac{1}{i} D^-_{ij}(x,y|\Phi_c),
\qquad
D^- = D^c - D^{adv},
\\
<T \hat{\varphi}_i(x|\Phi_c) \hat{\varphi}_j(y|\Phi_c)>
= \frac{1}{i} D^c_{ij}(x,y|\Phi_c),
\\
(:\hat{\varphi}_i(x|\Phi_c)\hat{\varphi}_j(y|\Phi_c):)^+
=
:\hat{\varphi}_i(x|\Phi_c)\hat{\varphi}_j(y|\Phi_c):
+ \frac{1}{i} \Delta_{ij}(x,y|\Phi_c),
\\
\Delta = D^c + D^{c*} - D^{ret} - D^{adv}.
\eey
Consider the current operator:
\beq
j_0^l(x|\Phi_c) = \frac{1}{i} W_0[\Phi_c]
\frac{\delta W_0[\Phi_c]}{\delta \Phi_c^l(x)}.
\l{4m.6}
\eeq
Then the properties of unitarity and causality of the operator
$W_0[\Phi_c]$ will have the form analogous to
\r{4u.3} and \r{4u.4}:
\bey
<0|j_0^l(y|v)|0> =
(<0|j_0^l(y|v)|0>)^*,
\\
\frac{\delta j_0^l(y|v)}{\delta \Phi^s(x)} = 0,
\qquad x \gsim y.
\eey
Variate the relation \r{4m.5} and make use of eq.\r{4m.4}. Analogously
to eq.\r{4u.6}, we obtain:
\beq
[\hat{\varphi}^j(x|\Phi_c),
\int j_0^k(y|\Phi_c) \delta \Phi_c^k(y) dy]
=
i \int D^{js}(x,y|\Phi_c)
(\delta K' \hat{\varphi})^s(y|\Phi_c) dy.
\l{4m.7}
\eeq
Property
\r{4m.7} specifies the operator
$j_0^k(y|\Phi_c)$ up to an additive constant.
It is fixed by a one-loop effective action
\r{4m.5a}:
\beq
\int dy j_0^k(y|\Phi_c) \delta \Phi_c^k(y)
= - \frac{1}{2} \int dy
:\hat{\varphi}^k(y|\Phi_c)
(\delta K' \hat{\varphi})^k (y|\Phi_c):
+ \delta A_1[\Phi_c].
\l{4m.8}
\eeq
The Poincare covariance,  unitarity and causality properties imply the
following relations on the one-loop effective action:
\beb
A_1[u_g\Phi_c] = A_1[\Phi_c],
\quad
Im \delta A_1[\Phi_c] = \frac{1}{4} Tr (\delta K' \Delta),
\\
\int dx dy \delta_1\Phi(x)
\frac{\delta^2A_1}{\delta \Phi_c(x) \delta \Phi_c(y)} \delta_2\Phi(y)
= \frac{1}{2i} Tr
[\delta_1K' D^c \delta_2K' D^c],
\\
supp \delta_1 \Phi < supp \delta_2 \Phi.
\l{4m.9}
\eeb
Formally, the solution of eqs.\r{4m.9} have the form:
\beq
A_1[\Phi_c] = \frac{i}{2} ln det [1+K'D_0^c];
\l{4m.10}
\eeq
definition of the determinant requires renormalization.

\section{Semiclassical perturbation theory}

Investigate the axioms of semiclassical field theory
C1-C5 within the perturbation framework.  It is convenient  to  fix  a
representation. Otherwise,  there  will  be  a  nonuniqueness  for the
operators $\underline{U}_g$,                     $\hat{\varphi}_h(x)$,
$\underline{\Phi}_R(x|J)$, $\underline{W}[\Phi_c]$.

\subsection{Asymptotic in-representation.   Axioms   of  semiclassical
S-matrix}

In quantum field theory,  the asymptotic  in-representation  is  often
used.  One  supposes  that at $t\to -\infty$ particles become free and
Hilbert state space coincides with the Fock space of  free  particles.
For such a case,  Heisenberg field $\hat{\varphi}_h(x)$ tends as $t\to
\pm \infty$ in a weak sense to the asymptotic free field:
\bez
\hat{\varphi}_h(x) \sim_{x^0\to -\infty} \hat{\varphi}_{in}(x)  \equiv
\hat{\varphi}_0(x),
\qquad
\hat{\varphi}_h(x)
\sim_{x^0\to +\infty} \hat{\varphi}_{out}(x).
\eez
The S-matrix
(denote it as $\Sigma[0]$)  is an unitary transformation of
asymptotic fields:
\beq
\hat{\varphi}_{out}(x) = \Sigma^+[0] \hat{\varphi}_{in}(x) \Sigma[0] =
\Sigma^+[0] \hat{\varphi}_{0}(x) \Sigma[0].
\l{5.1}
\eeq
The operator
\bez
\Sigma[\Phi_c] = \Sigma[0] \underline{W}[\Phi_c]
\eez
is an S-matrix in the external background.  Write  the  axioms  C1-C5,
making use of the introduced notations. As
$x^0\to -\infty$, formulas of the Poincare transformations take the form:
\beq
\underline{U}_{g_1g_2} = \underline{U}_{g_1} \underline{U}_{g_2},
\quad
\underline{U}_{g^{-1}} \hat{\varphi}_0(x)       \underline{U}_g      =
\hat{\varphi}_0(w_gx).
\l{5.2a}
\eeq
Since the asymptotic in-field is a free scalar field of the mass  $m$,
the   operator   $\underline{U}_g=U_g$  coincides  with  the  Poincare
transformation of the free theory.

Properties of  Poincare  invariance,  unitarity  and   causality   are
presented as:
\beb
U_g \Sigma[\Phi_c] U_{g^{-1}} = \Sigma[u_g\Phi_c],
\quad
(\Sigma[\Phi_c])^+ = (\Sigma[\Phi_c])^{-1},
\\
\frac{\delta}{\delta \Phi_c(x)}
j(y|\Phi_c) = 0,
\quad x \gsim y,
\l{5.2b}
\eeb
where $j(y|\Phi_c)$ is a current operator
\beq
j(y|\Phi_c) =
\frac{1}{i}
\Sigma^+[\Phi_c] \frac{\delta \Sigma[\Phi_c]}{\delta \Phi_c(y)},
\l{5.2c}
\eeq
entering to the right-hand side of the Yang-Feldman equation \r{3.12d}
\beq
(\partial_{\mu} \partial^{\mu} + V'{}'(\Phi_c(x)))
(\underline{\Phi}_R(x|J_{\Phi_c}) - \Phi_c(x)) = h j(x|\Phi_c).
\l{5.3}
\eeq
As $x^0\to        \pm        \infty$,        semiclassical       field
$\underline{\Phi}_R(x|J_{\Phi_c})$   will   satisfy   the   asymptotic
boundary conditions:
\beb
\underline{\Phi}_R(x|J_{\Phi_c}) \sim_{x^0\to    -\infty}
\sqrt{h} \hat{\varphi}_0(x),
\\
\underline{\Phi}_R(x|J_{\Phi_c}) \sim_{x^0\to    +\infty}
\underline{W}^+[\Phi_c]
\sqrt{h} \hat{\varphi}_{out}(x)
\underline{W}[\Phi_c]
=
\Sigma^+[\Phi_c]
\sqrt{h} \hat{\varphi}_0(x)
\Sigma[\Phi_c].
\l{5.4}
\eeb
The sollution of  eq.  \r{5.3}  satisfying  the  first  of  conditions
\r{5.4}  can  be  expressed  via  the  field  $\hat{\varphi}_v(x)$ and
retarded Green function:
\beq
\underline{\Phi}_R(x|J_{\Phi_c}) - \Phi_c(x) =
\sqrt{h} \hat{\varphi}_v(x) + h \int dy D_v^{ret}(x,y)
j(y|\Phi_c).
\l{5.5a}
\eeq
If we  use  the  second  condition,  the  semiclassical  field will be
expressed via the advanced Green function:
\beb
\underline{\Phi}_R(x|J_{\Phi_c}) - \Phi_c(x) =
\Sigma^+[\Phi_c]
W_0[\Phi_c]
\sqrt{h} \hat{\varphi}_v(x)
W_0^+[\Phi_c]
\Sigma[\Phi_c]
\\ + h \int dy D_v^{adv}(x,y)
j(y|\Phi_c).
\l{5.5b}
\eeb
Relations \r{5.5a} and \r{5.5b} imply the following identity:
\beq
[\hat{\varphi}_v(x),
W_0^+[\Phi_c] \Sigma[\Phi_c] ] =
\sqrt{h} \int dy D_v(x,y)
W_0^+[\Phi_c] \Sigma[\Phi_c] j(y|\Phi_c).
\l{5.6}
\eeq
Formula \r{5.6} can be viewed as a basis for the perturbation theory
for $\Sigma[\Phi_c]$:
\bez
\Sigma[\Phi_c] = \Sigma_0[\Phi_c] + \sqrt{h} \Sigma_1[\Phi_c] + ...
\eez
Namely, let   the   $k-1$-th  order  of  the  perturbation  theory  is
constructed.  Then $\Sigma_k[\Phi_c]$ is defined from the  commutation
relation    \r{5.6}    up    to    an    operator    of    the    form
$c_k[\Phi_c]W_0[\Phi_c]$, with a c-number multiplier $c_k$. To fix it,
introduce the  following  relation  for  the  vacuum  average  of  the
scattering operator in the external field:
\beq
<0|\underline{W}[\Phi_c]|0> =
<0|\Sigma[\Phi_c]|0> = e^{i(A_1[\Phi_c] + hA_2[\Phi_c] + ...)}.
\l{5a.0}
\eeq
Let us call the quantity
\bez
\underline{A}[\Phi_c] =  I[\Phi_c]  + hA_1[\Phi_c] + h^2A_2[\Phi_c] +
...
\eez
as a semiclassical action of the theory. It differs from the effective
action.  All  connected  Feynman  graphs  are  involved  to   it.   If
$\underline{A}[\Phi_c]$  is  given,  the  operator $\Sigma[\Phi_c]$ is
uniquely defined within the perturbation framework from  eq.  \r{5.6}.
Properties  of  Poincare  invariance,  unitarity  and causality impose
certain conditions on semiclassical action.  $A_k[\Phi_c]$ is  defined
up to a real local Poincare invariant functional.

Thus, we  are  coming  to  the following set of axioms of the S-matrix
approach.

{\bf S1}. {\it Hilbert state space
$\cal F$ is a Fock space for the free field,  Poincare transformations
coincide with free case.
}

{\bf S2.}  {\it
To each classical field configuration
$\Phi_c(x)$ with compact support one assigns S-matrix
$\Sigma[\Phi_c]$ being a formal asymptotic series in
$\sqrt{h}$. It   satisfies  the  Poincare  invariance,  unitariry  and
causality properties \r{5.2b}, as well as commutation relation \r{5.6}.
}

The retarded semiclassical field
$\underline{\Phi}_R(x|J)$ is specified from
$\Sigma[\Phi_c]$  uniquely, so that there are no additional conditions
on $\underline{\Phi}_R(x|J)$. Properties C3 should be obtained from S2.

Formula \r{5.5a}    specifies    the    classical    retarded    field
$\underline{\Phi}_R(x|J)$  as $J\sim 0$.  Making use of the Bogoliubov
causality  condition,  we  extend  the  definition  to  all  $J$.  The
properties   of   Hermitian   conjugation,   Poincare  invariance  and
Bogoliubov  causality  for  $\underline{\Phi}_R(x|J)$,  as   well   as
commutation relation \r{3.3c} can be checked.

\subsection{Perturbation theory}

Let us develop the perturbation theory for S-matrix  in  the  external
field (the first and the second order). Introduce the notation
\beq
\tilde{\Sigma}[\Phi_c]
= W_0^+[\Phi_c] \Sigma[\Phi_c]
= 1 + \sqrt{h}
\tilde{\Sigma}_1[\Phi_c] +
h \tilde{\Sigma}_2[\Phi_c] + ...
\l{5x.1}
\eeq
Then the current \r{5.2c},
with leading order of the form
\bez
j_0(y|\Phi_c) = \frac{1}{i} W_0^+[\Phi_c]
\frac{\delta W_0[\Phi_c]}{\delta  \Phi_c(y)}  =  V'{}'{}'  (\Phi_c(y))
\rho(y|v),
\eez
will be expressed via the operator \r{5x.1} as:
\beb
j(y|\Phi_c) =
\tilde{\Sigma}^+[\Phi_c] j_0(y|\Phi_c) \tilde{\Sigma}[\Phi_c]
+
\frac{1}{i}
\tilde{\Sigma}^+[\Phi_c] \frac{\delta   \tilde{\Sigma}[\Phi_c]}{\delta
\Phi_c(y)}
\\ =
j_0(y|\Phi_c) + \sqrt{h} j_1(y|\Phi_c) + ...
\l{5x.1a}
\eeb
Write the formulas for condtructing the perturbation theory:
\beb
[\hat{\varphi}_v(x),
\tilde{\Sigma}[\Phi_c]] =
\sqrt{h} \int dy D_v(x,y) \tilde{\Sigma}[\Phi_c] j(y|\Phi_c);
\\
<\tilde{\Sigma}[\Phi_c]> = e^{i(hA_2[\Phi_c] + h^2 A_3[\Phi_c] + ...)}
\l{5x.2}
\eeb

\subsubsection{First order}

In the first order, relations \r{5x.2} take the form
\beb
[\hat{\varphi}_v(x),
\tilde{\Sigma}_1[\Phi_c]] =
\sqrt{h} \int dy D_v(x,y) \tilde{\Sigma}[\Phi_c]
V'{}'{}'(\Phi_c(y)) \rho(y|v).
\\
<\tilde{\Sigma}_1[\Phi_c]> = 0.
\l{5x.4}
\eeb
To solve  these  equations,  it  is convenient to introduce
the following notations.
For the T-square, set
\beq
(T\hat{\varphi}_v^2(y))_R \equiv -2\rho(y|v) =
:\hat{\varphi}_v^2(y): + \frac{1}{i} (D_v^c(y,y))_R.
\l{5x.5}
\eeq
The operator
$\tilde{\Sigma}_1[\Phi_c]$ is rewritten as
\beq
\tilde{\Sigma}_1[\Phi_c] = - \frac{i}{6} \int  dy  V'{}'{}'(\Phi_c(y))
(T\hat{\varphi}_v^3(y))_R,
\l{5x.7}
\eeq
Eq.\r{5x.4} implies the following relation:
\beq
[\hat{\varphi}_v(x), (T\hat{\varphi}_v^3(y))_R] =
- 3i D_v(x,y) (T\hat{\varphi}_v^2(y))_R,
\quad
<(T\hat{\varphi}_v^3(y))_R> = 0,
\l{5x.8}
\eeq
The quantity $(T\hat{\varphi}_v^3(y))_R$ is uniquely defined.
\beq
(T\hat{\varphi}_v^3(y))_R =
:\hat{\varphi}_v^3(y): +
\frac{3}{i}
(D_v^c(y,y))_R \hat{\varphi}_v(y).
\l{5x.12}
\eeq
The Poincare invariance and unitarity properties are taken to the form:
\beq
U_g
(T\hat{\varphi}_v^3(w_gy))_R U_g^{-1}
= (T\hat{\varphi}_{u_gv}^3(y))_R;
\quad
(T\hat{\varphi}_v^3(y))_R^+ = (T\hat{\varphi}_v^3(y))_R.
\l{5x.9}
\eeq
To check these properties, it is sufficient to notice that
differences
$(T\hat{\varphi}_v^3(y))_R^+ - (T\hat{\varphi}_v^3(y))_R$
and
$U_g (T\hat{\varphi}_v^3(w_gy))_R U_g^{-1}
- (T\hat{\varphi}_{u_gv}^3(y))_R$
commute with
$\hat{\varphi}_v(x)$ and have zero matrix elements
$<...>$.

Investigate the causality condition. The current operator in the first
order of perturbation theory has the form:
\bey
j_1(y|\Phi_c) = - \frac{1}{6} V^{(IV)}(\Phi_c(y))
(T\hat{\varphi}_v^3(y))_R -
\\
\frac{1}{6}
\int dz
V'{}'{}'(\Phi_c(y)) V'{}'{}'(\Phi_c(z))
\left(
\frac{\delta}{\delta v(y)}
(T\hat{\varphi}_v^3(z))_R - \frac{i}{2}
[(T\hat{\varphi}_v^2(y))_R,
(T\hat{\varphi}_v^3(z))_R]
\right)
\eey
This operator     satisfies     the     causality     condition     if
$(T\hat{\varphi}_v^3(y))_R$ depends on $v(y)$ at the  preseeding  time
moments     anf    the    expression    $\frac{\delta}{\delta    v(y)}
(T\hat{\varphi}_v^3(z))_R  -  \frac{i}{2}  [(T\hat{\varphi}_v^2(y))_R,
(T\hat{\varphi}_v^3(z))_R]$ vanishes for $z \gsim y$.  Represent these
properties as
\bey
\frac{\delta}{\delta v(x)}
(T\hat{\varphi}_v^3(y))_R = 0, \quad x \gsim y,
\\
\frac{\delta}{\delta v(x)}
(T\hat{\varphi}_v^3(y))_R =
\frac{i}{2}
[(T\hat{\varphi}_v^2(x))_R,
(T\hat{\varphi}_v^3(y))_R], \quad x \lsim y.
\eey
These properties are checked analogously.

\subsubsection{The second order}

Write down eqs.\r{5x.2} for $\tilde{\Sigma}_2$:
\beb
[\tilde{\varphi}_v(x), \tilde{\Sigma}_2[\Phi_c]] =
\int dy D_v(x,y)
(j_1(y|\Phi_c) + \tilde{\Sigma}_1[\Phi_c] j_0(y|\Phi_c));
\\
<\tilde{\Sigma}_2[\Phi_c]> = iA_2[\Phi_c].
\l{5x.13}
\eeb
Use the explicit forms of
$j_0(y|\Phi_c))$, $j_1(y|\Phi_c)$, $\tilde{\Sigma}_1[\Phi_c]$ and take
eq.\r{5x.13} to the form:
\beb
\left[\tilde{\varphi}_v(x), \tilde{\Sigma}_2[\Phi_c] \right] =
-i \int dy D_v(x,y)
\left\{
- \frac{i}{6} V^{(IV)}(\Phi_c(y))
(T\hat{\varphi}_v^3(y))_R
\right.
\\
\left.
- \frac{1}{12} \int dz
V'{}'{}'(\Phi_c(y)) V'{}'{}'(\Phi_c(z))
(T\hat{\varphi}_v^2(y)\hat{\varphi}_v^3(z))_R
\right\}.
\l{5x.13a}
\eeb
We use the notation
\beq
(T\hat{\varphi}_v^2(y)\hat{\varphi}_v^3(z))_R
\equiv
(T \hat{\varphi}_v^2(y))_R
(T \hat{\varphi}_v^3(z))_R
+ 2i \frac{\delta}{\delta v(y)}
(T \hat{\varphi}_v^3(z))_R.
\l{5x.15}
\eeq
The renormalized    T-product    \r{5x.15}    coinsides    with    $(T
\hat{\varphi}_v^2(y))_R (T \hat{\varphi}_v^3(z))_R$ as $y\gsim z$, and
with $(T \hat{\varphi}_v^3(z))_R(T \hat{\varphi}_v^2(y))_R$ as $z\gsim
y$, so that notation \r{5x.15} is reasonable.

The solution of eq.\r{5x.13a} can be formally presented as
\beb
\tilde{\Sigma}_2[\Phi_c] =
- \frac{i}{24} \int dy V^{(IV)}(\Phi_c(y))
(T \hat{\varphi}_v^4(y))_R
\\
-
\frac{1}{72} \int dy dz
V'{}'{}'(\Phi_c(y)) V'{}'{}'(\Phi_c(z))
(T\hat{\varphi}_v^3(y)\hat{\varphi}_v^3(z))_R.
\l{5x.13b}
\eeb
New renormalized  T-products  entering to eq.\r{5x.13b} should satisfy
the commutation relations:
\beb
\left[\hat{\varphi}_v(x),
(T\hat{\varphi}_v^4(y))_R
\right] =
\frac{4}{i} D_v(x,y)
(T\hat{\varphi}_v^3(y))_R,
\\
\left[\hat{\varphi}_v(x),
(T\hat{\varphi}_v^3(y)\hat{\varphi}_v^3(z))_R
\right] = \\
\frac{3}{i} D_v(x,y)
(T\hat{\varphi}_v^2(y)\hat{\varphi}_v^3(z))_R
+
\frac{3}{i} D_v(x,z)
(T\hat{\varphi}_v^3(y)\hat{\varphi}_v^2(z))_R.
\l{5x.14}
\eeb
The renormalized T-products are specified from eq.
\r{5x.14} up to a c-number funcitons
$<...>$. They are related with two-loop semiclassical action:
\bey
A_2[\Phi_c] =
- \frac{1}{24} \int dy V^{(IV)}(\Phi_c(y))
<(T \hat{\varphi}_v^4(y))_R>
+ \\
\frac{i}{72} \int dy dz
V'{}'{}'(\Phi_c(y)) V'{}'{}'(\Phi_c(z))
<(T\hat{\varphi}_v^3(y)\hat{\varphi}_v^3(z))_R>.
\eey
Из \r{5x.14} вытекает, что:
\bey
(T\hat{\varphi}_v^4(y))_R =
:\hat{\varphi}_v^4(y): +
6 \frac{1}{i} (D_v^c(y,y))_R :\hat{\varphi}_v^2(y):
+ <(T\hat{\varphi}_v^4(y))_R>,
\\
(T\hat{\varphi}_v^3(y)\hat{\varphi}_v^3(z))_R =
:\hat{\varphi}_v^3(y)\hat{\varphi}_v^3(z): +
\frac{9}{i} D_v^c(y,z)
:\hat{\varphi}_v^2(y)\hat{\varphi}_v^2(z):
\\ +
\frac{3}{i} (D_v^c(z,z))_R
:\hat{\varphi}_v^3(y)\hat{\varphi}_v(z):
+ \frac{3}{i} (D_v^c(y,y))_R
:\hat{\varphi}_v(y)\hat{\varphi}_v^3(z):
\\ +
\frac{9}{i^2} D_v^c(y,z) (D_v^c(y,y))_R
:\hat{\varphi}_v^2(z): +
\frac{9}{i^2} D_v^c(y,z) (D_v^c(z,z))_R
:\hat{\varphi}_v^2(y):
\\ +
\frac{9}{i^2} (D_v^c(y,y))_R (D_v^c(z,z))_R
:\hat{\varphi}_v(y)\hat{\varphi}_v(z): +
\frac{9}{i^2} (D_v^c(y,z))_R^2
:\hat{\varphi}_v(y)\hat{\varphi}_v(z):
\\ +
<(T\hat{\varphi}_v^3(y)\hat{\varphi}_v^3(z))_R>.
\eey
Investigate the  corollaries  of  properties  of  Poincare invariance,
unitarity and  causality.  The  Poincare  invariance   and   unitarity
property imply analogously to \r{5x.9} that
\beb
U_g
(T\hat{\varphi}_v^4(w_gy))_R U_g^{-1}
= (T\hat{\varphi}_{u_gv}^4(y))_R;
\quad
(T\hat{\varphi}_v^4(y))_R^+ = (T\hat{\varphi}_v^4(y))_R;
\\
U_g
(T\hat{\varphi}_v^3(w_gy)\hat{\varphi}_v^3(w_gz))_R U_g^{-1}
= (T\hat{\varphi}_{u_gv}^3(y)\hat{\varphi}_{u_gv}^3(z))_R;
\\
(T\hat{\varphi}_v^3(y)\hat{\varphi}_v^3(z))_R^+
+ (T\hat{\varphi}_v^3(y)\hat{\varphi}_v^3(z))_R
\\ = (T\hat{\varphi}_v^3(y))_R
(T\hat{\varphi}_v^3(z))_R
+ (T\hat{\varphi}_v^3(z))_R
(T\hat{\varphi}_v^3(y))_R.
\l{5x.17}
\eeb
The causality property leads to the following relations on
$(T\hat{\varphi}_v^4(y))_R$:
\beb
\frac{\delta}{\delta v(x)}
(T\hat{\varphi}_v^4(y))_R = 0, \quad x \gsim y,
\\
\frac{\delta}{\delta v(x)}
(T\hat{\varphi}_v^4(y))_R =
\frac{i}{2}
[(T\hat{\varphi}_v^2(x))_R,
(T\hat{\varphi}_v^4(y))_R], \quad x \lsim y.
\l{5x.19}
\eeb
и $(T\hat{\varphi}_v^3(y)\hat{\varphi}_v^3(z))_R$:
\beb
(T\hat{\varphi}_v^3(y)\hat{\varphi}_v^3(z))_R
=
(T\hat{\varphi}_v^3(y))_R
(T\hat{\varphi}_v^3(z))_R,
\quad y \gsim z,
\\
\frac{\delta}{\delta v(x)}
(T\hat{\varphi}_v^3(y)\hat{\varphi}_v^3(z))_R = 0, \quad
x \gsim y \mbox{ и } x \gsim z,
\\
\frac{\delta}{\delta v(x)}
(T\hat{\varphi}_v^3(y)\hat{\varphi}_v^3(z))_R =
\frac{i}{2}
[(T\hat{\varphi}_v^2(x))_R,
(T\hat{\varphi}_v^3(y)\hat{\varphi}_v^3(z))_R], \quad
x \lsim y \mbox{ и } x \lsim z.
\l{5x.20}
\eeb
It follows from eq.
\r{5x.14} that  commutators  between left-hand and right-hand sides of
relations
\r{5x.17}, \r{5x.19}  and  \r{5x.20}
with field
$\hat{\varphi}_v(\xi)$ coincide;  it  is  sufficient then to check the
equalities for matrix elements
$<...>$. If we set
\bey
<(T\hat{\varphi}_v^4(y))_R> = 3
(\frac{1}{i}
((D_v^c(y,y))_R)^2,
\\
<(T\hat{\varphi}_v^3(y)\hat{\varphi}_v^3(z))_R> =
\frac{6}{i^3}
(D_v^c(y,z))^3_R +
\frac{9}{i^3} (D_v^c(y,y))_R (D_v^c(z,z))_R D_v^c(y,z),
\eey
the properties of Poincare invariance,  unitarity and  causality  will
be rewritten as:
\bey
(D_v^c(y,z))^3_R  = (D_v^c(y,z))^3, \qquad y\ne z,
\\
(D_v^c(w_gy,w_gz))^3_R = (D_v^c(y,z))^3_R,
\\
\frac{\delta}{\delta v(x)}
(D_v^c(y,z))^3_R
= -3 (D_v^c(y,z))^2_R D_v^c(x,y) D_v^c(x,z),
\\
2i Im
(D_v^c(y,z))^3_R =
(D_v^-(y,z))^3 + (D_v^-(z,y))^3
\\ - 3 (D_v^c(y,z))^{2*}_R \Delta_v(x,y)
+  3 (D_v^c(y,z))^{*}
(\Delta_v(x,y))^2
- (\Delta_v(x,y))^3.
\eey
Therefore, semiclassical  methods in the axiomatic field theory allows
us to construct the renormalized perturbation theory for the  S-matrix.
The obtained  results  is  in  agreement with Lagrangian (Hamiltonian)
field theory, since
\bez
\tilde{\Sigma}[\Phi_c] =
Texp \left(
- i \int dx \left[
\frac{\sqrt{h}}{6} V'{}'{}'(\Phi_c(x)) \hat{\varphi}_v^3(x)
+ \frac{h}{24} V^{(IV)}(\Phi_c(x)) \hat{\varphi}_v^4(x) + ...
\right]
\right)
\eez

\section{On semiclassical scalar electrodynamics}

Let us  generalize  the  obtained  results to the gauge theories.
Start from quantum electrodynamics.

\subsection{Scalar electrodynamics and its quatization}

Consider the scalar electrodynamics,  the model of field theory, which
consists of vector (electromagnetic) field
$A_{\mu}(x)$ interacting with the complex scalar field.
The Lagrangian of the theory has the form:
\beq
{\cal L} =
D_{\mu} \theta^*  D^{\mu}\theta  -  \frac{1}{h}  V(h\theta^*\theta)  -
\frac{1}{4} F_{\mu\nu} F^{\mu\nu},
\l{6.1}
\eeq
here $D_{\mu} = \partial_{\mu}  -  i\sqrt{h}A_{\mu}$  is  a  covariant
derivative,
$F_{\mu\nu}  =  \partial_{\mu}  A_{\nu}   -
\partial_{\nu} A_{\mu}$, $V$ is a potential of self-interaction of the
scalar field.  For the Hamiltonian theory,  the momenta conjugated  to
fields $A^k$, $\theta$, $\theta^*$ have the form:
\bez
E_k = F_{k0}, \qquad
\pi_{\theta} = D_0\theta,
\qquad
\pi_{\theta}^* = D_0\theta^*.
\eez
The momentum canonically conjugated to the field
$A_0$ is zero. Therefore, the considered system is a constrained system
\c{23}, the Hamiltonian density and constraints have the form:
\beb
{\cal H} =
\frac{1}{2} E_kE_k + \frac{1}{4} F_{ij}F_{ij}
+ \pi_{\theta}^* \pi_{\theta} +
D_i\theta^* D_i\theta +
\frac{1}{h} V(h\theta^*\theta),
\\
\Lambda =  \partial_kE_k  +   i\sqrt{h}   (\pi_{\theta}^*   \theta   -
\pi_{\theta} \theta^*).
\l{6.1a}
\eeb
To quantize the theory,  one should substitute the fields and  momenta
by operators satisfying the canonical commutation relation.
There are different approaches to take the constraints into account.

For the Dirac approach, state
$\Psi_D$ is supposed to satisfy not only Schrodinger equation but also
additional constrained condition:
\bez
\hat{\Lambda}_{\bf x} \Psi_D = 0.
\eez
Since the constrained operators commute each other for this model, the
Dirac condition does not contradict Schrodinger equation.

The main  difficulty  of  the  Dirac approach is to introduce an inner
product. To avoid it, one can introduce an additional gauge condition,
for example $\partial_kA^k=0$ \c{2,23,24}.

Another way  to  quantize  constrained  systems  is to use the refined
algebraic quantization approach
\c{25}. There are no additional conditions on
state vectors $\Psi_A$;  instead,  the inner product of the theory  is
modified:
\bez
(\Psi_A,\Psi_A)_A = (\Psi_A,\prod_{\bf x} \delta(\hat{\Lambda}_{\bf x}
\Psi_A).
\eez
Two state vectors are set to be equivalent if their difference  is  of
zero norm.  In particular, state of the form $\int d{\bf x} \beta({\bf
x}) \Lambda_{\bf x} Y \sim 0$ is equivalent to zero.

State vectors $\Psi_D$ and $\Psi_A$ are related to each other  by  the
relation:
\bez
\Psi_D = \prod_{\bf x} \delta(\hat{\Lambda}_{\bf x}) \Psi_A.
\eez

A manifestly covariant approach to quantize gauge fields is a BRST-BFV
quantization approach
\c{26,27,28}.
In this approach,  additional degrees of freedom should be introduced.
These are Lagrange multipliers and Faddeev-Popov ghosts and antighosts
\c{2}  and canonically conjugated momenta.
When the Abelian case  is  investigated,  one  can  use  the  Lagrange
multipliers
$A_0({\bf  x})$   and momenta
$E_0({\bf x})$.  This  is a Gupta-Bleuler approach.  If the functional
Schrodinger representation is used, states
$\Psi_B$ are functionals of the fields
$A^k({\bf   x})$,   $A^0({\bf   x})$,
$\theta({\bf x})$, $\theta^*({\bf x})$:
\beq
\Psi_B = \Psi_B[A^k,A^0,\theta,\theta^*],
\l{6.3}
\eeq
while fields and momenta are operators
\bey
\hat{A}^{\mu}({\bf x}) = A^{\mu}({\bf x}),\quad
\hat{\theta}({\bf x}) = \theta({\bf x}), \quad
\hat{\theta}^*({\bf x}) = \theta^*({\bf x}),
\\
\hat{E}_{\mu}({\bf x}) =
\frac{1}{i} \frac{\delta}{\delta A^{\mu}({\bf x)}},
\quad
\hat{\pi}_{\theta}({\bf x}) =
\frac{1}{i} \frac{\delta}{\delta \theta^{*}({\bf x)}},
\quad
\hat{\pi}^*_{\theta}({\bf x}) =
\frac{1}{i} \frac{\delta}{\delta \theta({\bf x)}}.
\eey
An indefinite  inner  product  may  be  presented  via  the functional
integral of the form \c{29}
\beq
(\Psi_B,\Psi_B)_B =
\int DA^k D\lambda D\theta D\theta^*
(\Psi_B[A^k,-i\lambda,\theta,\theta^*])^*
\Psi_B[A^k,-i\lambda,\theta,\theta^*].
\l{6.4}
\eeq
All functionals \r{6.3} form a space of virtual states, while physical
states should obey the Gupta-Bleuler condition
\beq
\left[
\frac{1}{i} \frac{\delta}{\delta A^0({\bf x})} -
\frac{i}{\sqrt{-\Delta}} \Lambda_{\bf x}
\right] \Psi_B = 0.
\l{6.5}
\eeq
Two physical states with the difference of the form
\beq
\int d{\bf x} \beta({\bf x})
\left[
\frac{1}{i} \frac{\delta}{\delta A^0({\bf x})} +
\frac{i}{\sqrt{-\Delta}} \Lambda_{\bf x}
\right] Y_B \sim 0.
\l{6.6}
\eeq
are set to be equivalent.  "Zero" states \r{6.6} are orthogonal to all
physical  states.  Note  that  Gupta-Bleuler condition and equivalence
relation are invariant under evolution.

It follows from eq.\r{6.5} that
\bez
\Psi_B[A^k,A^0,\theta,\theta^*] =
e^{-\int d{\bf     x}     A^0({\bf     x})    \frac{1}{\sqrt{-\Delta}}
\hat{\Lambda}_{\bf x}} \Psi_A[A^k,\theta,\theta^*],
\eez
the inner product \r{6.4} is taken to the form $(\Psi_A,\prod_{\bf  x}
\delta(\hat{\Lambda}_{\bf  x}  \Psi_A)$,  so  that  the  Gupta-Bleuler
approach is equivalent to the algebraic quantization.

When the $\xi$-gauge is used,  the following terms are  added  to  the
Hamiltonian density:
\beq
{\cal H}_B = {\cal H} + A_0 \Lambda_{\bf x}
- \frac{\xi}{2} E_0^2 - E_0 \partial_k A^k.
\l{6.7}
\eeq
It follows  from  eqs.\r{6.5} and \r{6.6} that ${\cal H}_B \Psi_B \sim
{\cal H} \Psi_B$,  so that theories with  Haniltonians  $\cal  H$  and
${\cal H}_B$ are equivalent.

Starting from  the  Hamiltonian  \r{6.7},  one  obtains  the following
Lagrangian:
\beq
{\cal L}_B =
D_{\mu} \theta^* D^{\mu}\theta - \frac{1}{h} V(h\theta^*\theta) -
\frac{1}{4} F_{\mu\nu}F^{\mu\nu}  -   \frac{1}{2\xi}   (\partial_{\mu}
A^{\mu})^2.
\l{6.8}
\eeq
An additional gauge-fixed term is added.  Equations of motion for  the
model \r{6.7} (or \r{6.8}) are the following:
\bey
D_{\mu}D^{\mu} \theta + V'(h\theta^*\theta) \theta = 0,
\quad
E_0 = - \frac{1}{\xi} \partial_{\alpha} A^{\alpha},
\\
\partial_{\nu} F^{\mu\nu} + \partial^{\mu} E^0 = i\sqrt{h}
(\theta^* D^{\mu}\theta - D^{\mu}\theta^* \theta).
\eey
This implies that
$E_0(x)$ is  a  scalar  field  satisfying  the  massless  Kelin-Gordon
equation:
\beq
\partial_{\mu}\partial^{\mu} E_0 = 0,
\l{6.9}
\eeq
the constraint  $\Lambda$  is  related  with  the  field  $E_0$ by the
relation  $\dot{E}_0  =  -  \Lambda$.  Therefore,  the  quantum  field
$\hat{E}_0(x)$  can  be  expanded  to positive- and negative-frequency
components
\bez
\hat{E}_0(x) = \hat{E}_0^+(x) + \hat{E}_0^-(x).
\eez
The Gupta-Bleuler condition \r{6.5} is presented as
\beq
\hat{E}_0^-(x) \Psi_B = 0.
\l{6.10}
\eeq

\subsection{Additional axioms for the
semiclassical electrodynamics}

Let us   investigate   specific   features   of  semiclassical  scalar
electrosynamics.  As semiclassical virtual  states,  we  consider  the
vectors \r{2.2a} and their superpositions \r{2.4}.  Let $\Phi({\bf x})
\equiv ({\cal A}^{\mu}({\bf x}),  \Theta({\bf x}),\Theta^*({\bf  x}))$
be set of classical fields,  $\Pi({\bf x}) \equiv ({\cal E}_{\mu}({\bf
x}),  \Pi_{\theta}({\bf x}),\Pi_{\theta}^*({\bf x}))$  be  canonically
conjugated momenta.

Not all semiclassical states are physical.  Namely,  the Gupta-Bleuler
condition \r{6.10} leads to the  nontrivial  relations  for  classical
variables $X$ and vector $f\in {\cal F}_X$.

Analogously to \r{2.5} and \r{2.6},  for the operator
$\hat{E}_0(x)$ one can write
\bez
\sqrt{h} \hat{E}_0(x) K^h_X f \simeq K^h_X \underline{\cal E}_0  (x|X)
f.
\quad
\underline{\cal E}_0(x|X) =
{\cal E}_0(x|X) + \sqrt{h} {\cal E}_0^{(1)}(x|X) + ...,
\eez
with c-number function ${\cal E}_0(x|X)$. Physical states will satisfy
the conditions:
\beq
{\cal E}_0(x|X) = 0, \qquad
\underline{\cal E}_0^-(x|X) f =0,
\l{6.11}
\eeq
states of the form
\beq
f = \int dx \beta(x) \underline{\cal E}_0^+(x|X) g \sim 0
\l{6.12}
\eeq
are equivalent to zero. Conditions ${\cal  E}_0(x|X)=0$  and  $\dot{\cal
E}_0(x|X) = 0$ mean that
\beq
{\cal E}_0 = 0, \quad
\partial_k {\cal  E}_k  +  i(\Pi_{\theta}^*  \Theta   -   \Pi_{\theta}
\Theta^*) = 0.
\l{6.11a}
\eeq

Write analogs of the properties
\r{2.6v},   \r{2.13n}   and   \r{2.16}   for the semiclassical field
$\underline{\cal E}_0(x|X)$:
\beb
\underline{\cal E}_0(x|u_gX)
\underline{U}_g(u_gX \gets X) =
\underline{U}_g(u_gX \gets X) \underline{\cal E}_0(w_gx|X);
\\
ih \frac{\partial}{\partial \alpha} \underline{\cal E}_0(x|X) =
[\underline{\cal E}_0(x|X),        \underline{\omega}_X[\frac{\partial
X}{\partial \alpha}],
\\
\underline{\cal E}_0(x|X_2) \underline{V}(X_2\gets X_1)
= \underline{V}(X_2\gets X_1) \underline{\cal E}_0(x|X_1),
\quad
X_1 \sim X_2.
\l{6.13}
\eeb
The relations
\r{6.13} leads to the following important properties of  semiclassical
transformations:

(a) semiclassical Poincare transformation
$u_g$ takes physical state to physical;
semiclassical transformation
$\underline{U}_g(u_gX \gets   X)$
conserves the condition for physical states and equivalence relation.

(b) the     operator     $ih\frac{\partial}{\partial     \alpha}     +
\underline{\omega}_X[\frac{\partial   X}{\partial   \alpha}]$    takes
physical   semiclassical   states   to   physical  and  conserves  the
equivalence relation (this property means that for  $\alpha$-dependent
physical      state      $K^h_{X(\alpha)}f(\alpha)$      state     $ih
\frac{\partial}{\partial  \alpha}  K^h_{X(\alpha)}f(\alpha)$  is  also
physical);

(c) the  operator  $\underline{V}(X_2\gets X_1)$ takes physical states
to physical and conserves the equivalence property.

In semiclassical mechanics of constrained systems  \c{30},  there  are
also    equivalent    states    corresponding   to   different   field
configurations. Consider the operator:
\beq
e^{\frac{i}{\sqrt{h}}\int d{\bf x}  \gamma_1({\bf  x})  \hat{E}_0({\bf
x})},
\quad
e^{\frac{i}{\sqrt{h}}\int d{\bf x}  \gamma_1({\bf  x})
\hat{\Lambda}({\bf x})},
\l{6.14}
\eeq
with real  functions $\gamma_{1,2}({\bf x})$.  It follows from \r{6.6}
that each operator \r{6.14} takes a physical state to equivalent  one.
The classical configuration is transformed as:
\bey
{\cal A}^0 \to {\cal A}^0 - \gamma_1,\quad
{\cal A}^k \to {\cal A}^k + \partial_k \gamma_2,
\\
\Theta \to \Theta e^{i\gamma_2}, \quad
\Pi_{\theta} \to \Pi_{\theta} e^{i\gamma_2}.
\eey
Therefore, there  is  a  gauge  (physical)  equivalence  of  classical
states;  for  each  pair  of   gauge   equivalent   classical   states
$X_1\sim_{phys}  X_2$ an unitary operator $\underline{V}_{ph}(X_2\gets
X_1):  {\cal F}_{X_1} \to {\cal F}_{X_2}$ is specified.  It  satisfies
the relations analogous to \r{2.14}, \r{2.15} и \r{2.17}.

Therefore, the axioms of semiclassical field theory should be modified
for QED as follows.

{\bf G1 (QED)} {\it
A virtual  semiclassical  bundle  is  given,  space  of  the bundle is
interpreted as a set of virtual semiclassical states, base
${\cal X} = \{X\}$ is a set of virtual classical states,
fibres ${\cal F}_X$ (indefinite inner product spaces)
are spaces of quantum states in the external background.
}

In axioms G2-G5,  semiclassical bundle should be viewed as  a  virtual
semiclassical    bundle,    field    $\underline{\Phi}$   is   a   set
$(\underline{\cal  A}^{\mu},\underline{\Theta},\underline{\Theta}^*)$,
property   \r{2.6v}   should   be   modified   for  the  vector  field
$\underline{\cal A}^{\mu}$.

In addition, one should postulate the properties of the field
$\underline{\cal E}_0$.

{\bf G6 (QED)}. {\it
An operator-valued  distribution
$\underline{\cal  E}_0(x|X)$ expanded in
$\sqrt{h}$  is given. The leading order of expansion
${\cal E}_0(x|X)$ is a c-number. The following condition
is satisfied:
\beq
\partial_{\mu} \partial^{\mu} \underline{\cal E}_0(x|X) = 0
\l{6.14a}
\eeq
Properties \r{6.13}  are  obeyed.  Elements $X \in {\cal X}$ such that
${\cal E}_0(x|X) =0$ forms a set  of  all  classical  physcial  states
${\cal X}_{phys}$.  Elements of the space of the virtual semiclassical
bundle  such  that  $X\in  {\cal   X}_{phys}$   and   $\underline{\cal
E}_0^-(x|X)  f  =  0$ are physical semiclassical states.  Two physical
states $f_1,f_2\in {\cal F}_X$ are equivalent  iff  for  some  $g  \in
{\cal F}_X$ the difference $f_1-f_2$ has the form \r{6.12}. The scalar
square of physical semiclassical state $(X,f)$ is nonzero; it vanishes
iff $f\sim 0$. }

Take into account the property of gauge invariance.

{\bf G7 (QED)}. {\it
A gauge equivalence relation is specified on
${\cal  X}_{phys}$. For each pair
$X_1 \sim_{phys} X_2$ an unitary operator
$\underline{V}_{ph}:  {\cal  F}_{X_1}  \to  {\cal  F}_{X_2}$
taking physical states to physical, equivalent states to equivalent.
The following relations are satisfied:

- for $X_1 \sim X_2$, the properties
$X_1 \simp X_2$ and
$\underline{V}_{ph}(X_2 \gets X_1) \simp \underline{V}(X_2 \gets X_1)$
are satisfied;

- for $X_1 \simp X_2$ and $X_2 \simp X_3$, one has
$X_1\simp X_3$ и $\underline{V}_{ph}(X_3\gets X_1) \simp
\underline{V}_{ph}(X_3\gets X_2)\underline{V}_{ph}(X_2\gets X_1)$;

- let $X_1 \simp X_2$, then
$u_gX_1\simp u_gX_2$ и $\underline{V}_{ph}(u_gX_2 \gets u_gX_1)
\underline{U}_g(u_gX_1 \gets X_1)
\simp
\underline{U}_g(u_gX_2 \gets X_2)
\underline{V}_{ph}(X_2 \gets X_1)$;

- let $X_1(\alpha) \simp X_2(\alpha)$, then
$ih \frac{\partial}{\partial \alpha}
\underline{V}_{ph} (X_2 \gets X_1) \simp
\underline{V}_{ph} (X_2 \gets X_1)
\underline{\omega}_{X_1} [\frac{\partial X_1}{\partial \alpha}] -
\underline{\omega}_{X_2} [\frac{\partial X_2}{\partial \alpha}]
\underline{V}_{ph} (X_2 \gets X_1)$.
}

\subsection{Specific features    of    covariant    formulation     of
semiclassical electrodynamics}

Let us  discuss  the  axioms of covariant formulation of semiclassical
electrodynamics. Let us construct an analog of the state
\r{2.3}:
\beq
\Psi \simeq
e^{\frac{i}{h}\overline{S}}
Te^{\frac{i}{\sqrt{h}} \int dx [J_{\mu}(x) \hat{A}^{\mu}_h(x) +
\sigma(x) \hat{\theta}^*_h(x) +
\sigma^*(x) \hat{\theta}_h(x) +
\kappa(x) \hat{E}_{0h}(x)]} \overline{f}
\equiv
e^{\frac{i}{h} \overline{S}} T_J^h \overline{f}.
\l{6.16}
\eeq
Here $J  \equiv  (J_{\mu},\sigma^*,\sigma,\kappa)$  is  a  set of real
functions,  classical Schwinger sources.  In addition to  the  sources
$A^{\mu}$ and $\theta$, the source for the field $E_0$ is introduced.

Semiclassical retarded fields
\bez
\underline{\Phi}_R \equiv
(\underline{\cal A}^{\mu}_R,
\underline{\Theta}_R,
\underline{\Theta}_R^*,
\underline{\cal E}_{0R})
\eez
are introduced analogously to \r{3.1}.
Analogously to
\c{20}, we show that classical retarded field vanishes at
$-\infty$ and satisfy the system of equation
\beb
\partial_{\nu} {\cal F}^{\mu\nu}_R +
\partial^{\mu} {\cal E}_{0R} =
- i ({\cal D}_R^{\mu} \Theta_R^* \Theta_R -
\Theta_R^* {\cal D}_R^{\mu} \Theta_R) + J^{\mu},
\\
{\cal D}_{\mu R} {\cal D}^{\mu}_R + V'(\Theta^*_R \Theta_R) \Theta_R =
\sigma,
\quad
\partial_{\mu} {\cal A}^{\mu}_R + \xi {\cal E}_{0R} = - \kappa.
\l{6.17}
\eeb
Here the notations
${\cal  F}_{\mu\nu  R}  =
\partial_{\mu} {\cal A}_{\nu R} -
\partial_{\nu} {\cal A}_{\mu R}$, ${\cal D}_R^{\mu} = \partial^{\mu} -
i {\cal A}^{\mu}_R$ are used.
Properties of semiclassical retarded fields
\r{3.3o},  \r{3.3a},
\r{3.3b}, \r{3.3c},   \r{3.3d}
are generalized  for  electrodynamics;  when  the  Poincare invariance
property is written, one should take into account that
$\underline{\cal A}^{\mu}_R$ is a vector field.

The operator   $\underline{W}_J$,   functional   $\overline{I}_J$  and
equivalence relation $J\sim 0$ are  defined  according  to  eq.\r{3.5}
which is  viewed  in virtual sense.  It happens that $J\sim 0$ iff the
classical retarded field generated by $J$ is zero at $+\infty$. System
\r{6.17}    gives    a   one-to-one   correspondence   between   field
configurations            $\Phi_c            \equiv             ({\cal
A}_c^{\mu},\Theta_c,\Theta_c^*,{\cal  E}_{0c})$  with  compact support
and sources equivalent to zero.

Classical action is defined from relation \r{3.10}. It happens that it
is presented as a sum of a gauge-invariant part and gauge-fixing term:
\beb
I[\Phi_c] = I_{inv}[{\cal A}^{\mu}_c,\Theta_c,\Theta_c^*] +
I_{gf}[{\cal E}_{0c},{\cal A}^{\mu}_c],
\\
I_{inv} =   \int   dx   [-\frac{1}{4}   {\cal   F}_{\mu\nu   c}  {\cal
F}^{\mu\nu}_c + {\cal D}_{\mu c} \Theta_c^* {\cal D}^{\mu}_c  \Theta_c
- V(\Theta_c^* \Theta_c)],
\\
I_{gf} = \int dx [{\cal E}_{0c} \partial_{\mu} {\cal A}^{\mu}_c +
\frac{\xi}{2} {\cal E}_{0c}^2],
\l{6.18}
\eeb
with
${\cal D}_{\mu c} = \partial_{\mu} - i{\cal A}_{\mu c}$,
${\cal F}^{\mu\nu}_c    =    \partial^{\mu}    {\cal    A}^{\nu}_c   -
\partial^{\nu} {\cal A}^{\mu}_c$.
The properties of the operator
$\underline{W}[\Phi_c]$ are also generalized to electrodynamics.

Thus, axioms C1-C5 are modified as follows:
$\cal F$ is a space with indefinite inner product, Poincare invariance
should be written for vector case.

Investigate now  what additional conditions should be imposed in order
to obtain properties G6 and G7.

Relation \r{6.14a} can be rewritten as
\beq
\partial_{\mu} \partial^{\mu} \underline{\cal E}_{0R}(x|J) = 0,
\quad x \gsim supp J.
\l{6.20}
\eeq
Classical source $J$  corresponds to the classical physical state
iff
\beq
{\cal E}_{0R}(x|J) = 0, \quad x\gsim supp J.
\l{6.21a}
\eeq
The Gupta-Bleuler condition can be rewritten as
\beq
{\cal E}_{0R}^-(x|J) \overline{f} = 0.
\l{6.21}
\eeq

Investigate now gauge equivalence of semiclassical states. Notice that
the operators \r{6.14}  can be viewed as limit cases of the
operator
\beq
T e^{\frac{i}{\sqrt{h}} \int dx \hat{E}_0(x) \Delta \kappa(x)}
\l{6.22}
\eeq
The operator
\r{6.22} takes  physical  states   to   equivalent.   Therefore,   the
semiclassical states of the form
\bez
T^h_{J+\Delta \kappa_1} \overline{f}
\sim_{phys}
T^h_{J+\Delta \kappa_2} \overline{f},
supp \kappa_{1,2} > supp J
\eez
are physically    equivalent.    Here    $J+\Delta    \kappa    \equiv
(J_{\mu},\sigma^*,\sigma,\kappa+\Delta  \kappa)$.  Note that classical
retarded   field   generated   by   souces   $J+\Delta\kappa_1$    and
$J+\Delta\kappa_2$  are  gauge equivalent,  since they coincide in the
domain $supp J<x< supp \Delta \kappa_{1,2}$,  while for the $x>supp J$
case  the  function $\Delta \kappa$ enters to the gauge condition only
$\partial_{\mu} {\cal A}^{\mu} = - \Delta \kappa$.

We say that physical sources $J_1$ and  $J_2$  are  gauge  equivalent,
$J_1\sim_{phys} J_2$, iff any of the following properties is obeyed:

- there exist sources $\Delta \kappa_{1,2}$ satisfying the conditions
$supp \Delta \kappa_{1,2} > supp J_{1,2}$,  $J_1+\Delta \kappa_1  \sim
J_2 + \Delta \kappa_2$;

- there  exist  sources  $\Delta \kappa_{1,2}$ и $J_+$ such that $supp
J_+ > supp \Delta \kappa_{1,2} > supp J_{1,2}$,  $J_1+\Delta  \kappa_1
+J_+ \sim 0$, $J_2 + \Delta \kappa_2 + J_+ \sim 0$;

- the  retarded  fields  generated  by the sources $J_1$ and $J_2$ are
gauge equivalent.

We say that  $(\overline{S}_1,J_1)  \sim_{phys}  (\overline{S}_2,J_2)$
iff
\bez
\overline{S}_1 + \overline{I}_{J_1+\Delta\kappa_1 + J_+} =
\overline{S}_2 + \overline{I}_{J_2+\Delta\kappa_2+ J_+}.
\eez
Since ${\cal  E}_{0R}(x|J_1)=  {\cal  E}_{0R}(x|J_2)$  in  the  domain
$x>supp J_{1,2}$,  while $I_{inv}$ is gauge invariant, this definition
does not depend on the particular choice $\Delta \kappa_{1,2}$.

Set
\beq
\underline{V}_{ph}(X_2\gets X_1) =  \underline{V}(J_2+\Delta  \kappa_2
\gets J_1 + \Delta \kappa_1).
\l{6.25}
\eeq
If the property
\beq
\underline{\cal E}_{0R}(x|J+\Delta \kappa) =
\underline{\cal E}_{0R}(x|J),
\quad
supp \Delta \kappa > supp J,
\l{6.24}
\eeq
is satisfied for any physical source $J$,  the properties G7  will  be
satisfied. Namely,
\bez
\underline{\cal E}_{0R}(x|J_2)
\underline{V}_{ph}(X_2\gets X_1) =
\underline{V}_{ph}(X_2\gets X_1)
\underline{\cal E}_{0R}(x|J_1),
\quad
x > supp J_{1,2}.
\eez
Therefore, the  operator  $\underline{V}_{ph}(X_2\gets   X_1)$   takes
physical states to physical and conserves the equivalence property.

Show that the definition \r{6.25} does not depend on particular choice
of $\Delta \kappa_{1,2}$.
Under small variations, one has:
\bey
ih \delta \underline{V}_{ph}(X_2\gets X_1) =
\\
- \underline{V}_{ph}(X_2\gets X_1)
\int dx \underline{\cal E}_{0R} (x|J_1+\Delta \kappa_1)
\delta \Delta \kappa_1(x)
\\
+
\int dx \underline{\cal E}_{0R} (x|J_2+\Delta \kappa_2)
\delta \Delta \kappa_2(x)
\underline{V}_{ph}(X_2\gets X_1)
\\
=
- \underline{V}_{ph}(X_2\gets X_1)
\int dx \underline{\cal E}_{0R} (x|J_1)
\delta \Delta \kappa_1(x)
+
\int dx \underline{\cal E}_{0R} (x|J_2)
\delta \Delta \kappa_2(x)
\underline{V}_{ph}(X_2\gets X_1)
\eey
Therefore,
$\delta \underline{V}_{ph}(X_2\gets X_1)$ is zero  up  to  equivalence
relation. Other properties G7 are also checked by a direct calculation.

Notice that properties
\r{6.20}  and \r{6.24}  can be reduced to the one equation
\beq
\partial_{\mu} \partial^{\mu}    \underline{\cal    E}_{0R}(x|J+\Delta
\kappa) = 0, \quad x> supp J, supp\Delta \kappa > supp J.
\l{6.26}
\eeq
Namely, relation \r{6.20} is a partial case of \r{6.26}.  Moreover, it
follows   from   eqs.  \r{6.26}  and  \r{6.20}  that  $\underline{\cal
E}_{0R}(x|J)$    and    $\underline{\cal    E}_{0R}(x|J+\Delta\kappa)$
satisfy the same equation and coincide at
$supp J < x < supp \Delta \kappa$. Thus, we obtain property
\r{6.24}.

Thus, in addition to properties C1-C5,  the following axioms should be
satisfied.

{\bf C6 (QED)}. {\it The retarded semiclassical
field
$\underline{\cal E}_{0R}(x|J)$
satisfies the condition \r{6.26}.
Physical states satisfying the condition
\r{6.21}
has a nonnegative square of norm, which vanishes for the states
$\overline{f} = \int_{x\gsim  supp  J}  dx  \underline{\cal  E}_{0R}^+
(x|J) \beta(x) \overline{g}$ only.
}

Represent property   \r{6.26}   as   a   condition   on  the  operator
$W[\Phi_c]$.  In the domain $x >  supp  J$,  the  field  $\Phi_c(x)  =
\Phi_R(x|J+\Delta\kappa)$ satisfies the equations
\beq
\partial_{\nu} {\cal F}^{\mu\nu}_c +
i ({\cal D}^{\mu}_c \Theta_c^* \Theta_c
- \Theta_c^* {\cal D}^{\mu}_c \Theta_c)  = 0,
\\
{\cal D}_{\mu c} {\cal D}^{\mu}_c \Theta_c +
V'(\Theta_c^* \Theta_c) \Theta_c = 0.
\l{6.27}
\eeq
For functionals $F[\Phi_c]$, introduce the following notations:
\beb
\nabla_{\bf x} F \equiv
- \partial^{\mu}
\frac{\delta F}{\delta {\cal A}^{\mu}_c(x)}
+ i
\left[
\Theta_c(x) \frac{\delta F}{\delta \Theta_c(x)}
-
\Theta_c^*(x) \frac{\delta F}{\delta \Theta^*_c(x)}
\right],
\\
\Xi F \equiv \int dy \frac{\delta F}{\delta \Phi_c(y)}
\left\{
\underline{\Phi}_R(y|J_{\Phi_c}) - \Phi_c(y)
\right\}
\l{6.28}
\eeb
The operator  $\nabla_{\bf  x}$ is related to the gauge transformation
of the functional $F$:
\bez
\nabla_{\bf x} F \equiv \frac{\delta}{\delta \alpha(x)}
F [{\cal A}_c^{\mu} + \partial^{\mu} \alpha,
\Theta_c e^{i\alpha},
\Theta_c^* e^{-i\alpha}];
\eez
if the functional $F$ is gauge invariant then $\nabla_{\bf x} F = 0$.

Under notations \r{6.28}, the Yang-Feldman relations
\r{3.12d} has the following form at the domain $x>supp J$:
\beb
\Xi \frac{\delta I_{inv}}{\delta {\cal A}_c^{\mu}(x)}
+
\partial_{\mu} \underline{\cal E}_{0R}(x|J+\Delta \kappa) =
ih
\underline{W}^+    \frac{\delta     \underline{W}}{\delta     {\cal
A}_c^{\mu}(x)},
\\
\Xi \frac{\delta I_{inv}}{\delta {\Theta}_c^{*}(x)}
= ih
\underline{W}^+    \frac{\delta     \underline{W}}{\delta
\Theta_c^{*}(x)},
\quad
\Xi \frac{\delta I_{inv}}{\delta {\Theta}_c(x)}
= ih
\underline{W}^+    \frac{\delta     \underline{W}}{\delta
\Theta_c(x)},
\l{6.29}
\eeb
Making use of relations \r{6.29}, we find:
\bez
\Xi \nabla_{\bf x} I_{inv}
- i
\left(
\Xi \Theta_c(x)
\frac{\delta I_{inv}}{\delta \Theta_c(x)}
-
\Xi \Theta_c^*(x)
\frac{\delta I_{inv}}{\delta \Theta_c^*(x)}
\right)
= ih \underline{W}^+ \nabla_{\bf x} \underline{W}.
\eez
Take into account that the functional $I_{inv}$ is gauge invariant and
$\nabla_{\bf  x}   I_{inv}   =   0$;   $\frac{\delta   I_{inv}}{\delta
\Theta_c(x)}=  0$ and $\frac{\delta I_{inv}}{\delta \Theta_c^*(x)}= 0$
for the  classical  field  at  $x>supp  J$.  Therefore,  the  relation
\r{6.26} means that the scattering matrix in the external field should
be gauge invariant:
\beq
\nabla_{\bf x} \underline{W} = 0,
\l{6.30}
\eeq
provided that classical equation of motion \r{6.27} are satisfied.

\subsection{On the leading order}

Fields and  S-matrix  $W_0$  are  constructed,  making  use of general
formulas of subsection 4.7.

For the free case, analogously to subsection 4.2 one obtains equations
and  commutation  relations  \r{4.1b}  for  multicomponent  free field
$\hat{\varphi}_0(x)       =        \hat{\varphi}_{in}(x)        \equiv
(\hat{A}^{\mu}_{in}(x),  \hat{\theta}_{in}(x), \hat{\theta}^*_{in}(x),
\hat{E}_0^{in}(x))$,    with     independent     components.     Field
$\hat{\theta}_{in}$  is  a  complex scalar free field of the mass $m$.
For electromagnetic field, we obtain:
\bey
\partial_{\nu} F_{in}^{\mu\nu} + \partial^{\mu} \hat{E}_0^{in} = 0,
\quad
\partial_{\mu} \hat{A}^{\mu}_{in} + \xi \hat{E}_0^{in} = 0,
\\
\left[\hat{A}^{\mu}_{in}(x), \hat{A}^{\nu}_{in}(y) \right]
= - \frac{1}{i}
\left(g^{\mu\nu} D_0(x-y) -
(1-\xi)
\right.
\\ \left. \times
\partial_x^{\mu} \partial_x^{\nu}
\int dz (D_0^{ret}(x-z)D_0^{ret}(z-y) -
D_0^{adv}(x-z)D_0^{adv}(z-y))
\right),
\\
\left[
\hat{A}^{\mu}_{in}(x), \hat{E}_0^{in}(y)] =
- \frac{1}{i} \partial_{\mu} D_0(x-y),
\quad
\left[
\hat{E}_0^{in}(x), \hat{E}_0^{in}(y) \right] = 0.
\right]
\eey
Here
$D_0$ is  a  commutation   function   for   massless   scalar   field.
Commutation relations  takes the simplest form at $\xi=1$;  it is this
gauge that is used usually in calculations.

Semiclassical field
$\hat{\varphi}(x|\Phi_c) \equiv
(\hat{A}^{\mu}(x),\hat{\theta}(x),\hat{\theta}^*(x),\hat{E}_0(x))$
satisfies the system \r{4m.1}. Write its explicit form for the case
$V(\Theta_c^*\Theta) = m^2 \Theta_c^* \Theta_c$:
\beb
\partial_{\nu} \hat{F}^{\mu\nu} +
\partial^{\mu} \hat{E}_0 +
i
(\widehat{{\cal D}^{\mu}_c \theta}^* \Theta_c
+ {\cal D}^{\mu}_c \Theta_c^* \hat{\theta}
- \hat{\theta}^* {\cal D}^{\mu}_c \Theta_c
- \Theta_c^* \widehat{{\cal D}^{\mu}_c \theta})
= 0,
\\
\partial_{\mu} \hat{A}^{\mu} + \hat{E}_0 = 0,
\quad {\cal D}_c^{\mu}
\widehat{{\cal D}^{\mu}_c \theta}
- i \hat{A}^{\mu} {\cal D}_{\mu c} \Theta_c = 0,
\l{6.30b}
\eeb
where
\bez
\widehat{{\cal D}^{\mu}_c \theta} =
\partial^{\mu} \hat{\theta}
- i \hat{A}^{\mu} \Theta_c -
i {\cal A}^{\mu}_c \hat{\theta}.
\eez

Components of current vector are
\bey
j_A^{\mu} \equiv
\frac{1}{i} W_0^+ \frac{\delta W_0}{\delta {\cal A}_{\mu}}
=
:i \hat{\theta}^*
\widehat{{\cal D}^{\mu}_c \theta}
-
i (\widehat{{\cal D}^{\mu}_c \theta})^*
\hat{\theta} + \hat{A}^{\mu}
(\Theta_c^* \hat{\theta} + \Theta_c \hat{\theta}^*):
+ \frac{\delta A_1}{\delta {\cal A}_{\mu c}};
\\
j_{\theta}^* \equiv
\frac{1}{i} W_0^+ \frac{\delta W_0}{\delta {\Theta}_{c}}
=
:- i \hat{A}_{\mu}
(\widehat{{\cal D}^{\mu}_c \theta})^*
- i {\cal D}^{\mu}_c
(\hat{A}_{\mu} \hat{\theta}^*):
+ \frac{\delta A_1}{\delta {\Theta}_{c}}.
\eey

Properties of Poincare invariance,  unitarity and  causality  lead  to
restrictions   \r{4m.9}   on  one-loop  effective  action.  The  gauge
invariant  property  \r{6.30}  leads  to  the  following  requirement.
$A_1[\Phi_c]$ should be gauge invariant:
\beq
\nabla_{\bf x} A_1[\Phi_c] = 0
\l{6.31}
\eeq
under conditions \r{6.27}.

The property of positive definiteness of  inner  product  in  physical
space is a corollary of general properties of indefinite inner product
spaces.

The further analysis of scalar electrodynamics  is  analogous  to  the
case  of  scalar  field.  For  each order of perturbation theory,  one
should additionally check property \r{6.30} which is  related  to  the
gauge invariance of classical action.

\section{Semiclassical integrals of motion and BRST-charge}

The quantizations considered  above are applicable only to Abelian theories.
A manifestly covariant approach to quantize nonablelian fields is BRST
approach.

\subsection{On BRST-quantization}

In BRST-quantization,  additional degrees of freedom are introduced to
the theory.  To each constraint $\Lambda^a$ one assigns  the  Lagrange
multiplier  $\lambda^a$,  canonically  conjugated momentum $\pi_a$ and
fermionic variables:  ghost $C^a$ and antighost  $\overline{C}_a$  and
canonically  conjugated  momenta  $\overline{\Pi}_a$ and $\Pi^a$.  The
operators $\overline{C}_a$ and $\overline{\Pi}_a$ are  anti-Hermitian,
others are Hermitian. The nontrivial commutation relations are:
\bez
[\lambda^a,\pi_b] = i\delta^a_b,
\quad
[C^a,\overline{\Pi}_b]_+ = \delta^a_b,
\quad
[\overline{C}_a,\Pi^b]_+ = \delta^b_a.
\eez
The main  object  of  the  theory  is  BRST-charge  $\hat{Q}$.  If the
constraints form a Lie algebra with structure constants $f^{abc}$
\bez
[\hat{\Lambda}_a, \hat{\Lambda}_b] = - i f^{abc} \hat{\Lambda}_c,
\eez
the BRST-charge has the following form:
\beq
\hat{Q} = C^a \hat{\Lambda}_a +
\frac{i}{2} f^{abc} \overline{\Pi}_a C^b C^c - i \pi_a \Pi^a.
\l{7c.1}
\eeq
It satisfies the Hermitian and nilpotent conditions:
\bez
\hat{Q}^+ = \hat{Q},
\quad \hat{Q}^2 = 0.
\eez
The ghost number operator is also introduced to the theory:
\bez
\hat{N} = \Pi^a \overline{C}_a - \overline{\Pi}_a C^a,
\eez
It is the difference between number of ghosts and antighosts.
It satisfies the commutation relations:
\bey
\left[\hat{N},C^a\right] = C^a, \quad \left[\hat{N},\Pi^a \right] = \Pi^a,
\\
\left[\hat{N},\overline{C}_a\right] = -\overline{C}_a,
\quad \left[\hat{N},\overline{\Pi}_a\right] = - \overline{\Pi}_a,
\quad
\left[\hat{N},\hat{Q}\right] = \hat{Q}.
\eey
The following conditions are imposed for physical states:
\beq
\hat{Q} \Psi_B = 0, \quad
\hat{N} \Psi_B = 0;
\l{7c.1b}
\eeq
physical states of the form
\bez
\hat{Q} Y_B \sim 0, \hat{N} Y_B = - Y_B,
\eez
being orthogonal to physical subspace are set to be equivalent to zero.

For the   scalar  electrodynamics,  the  BRST-quantization  method  is
equivalent to the Gupta-Bleuler  approach.  Namely,  for  system  with
Hamiltonian and constraints
\r{6.1a} the BRST-charge has the form:
\beq
\hat{Q} =  \int  d{\bf x} [C({\bf x}) \Lambda_{\bf x} - i E_0({\bf x})
\Pi({\bf x})].
\l{7c.2}
\eeq
The operator $[\hat{Q},K]_+$ being equivalent to zero is added to  the
Hamiltonian of the system. The operator $K$ is chosen to be as follows:
\beq
K = \int d{\bf x}
[A_0({\bf x}) \overline{\Pi}({\bf x})
- i \overline{C}({\bf x}) \partial_k A^k({\bf x})
- \frac{i}{2} \xi \overline{C}({\bf x}) E_0({\bf x})].
\l{7c.2a}
\eeq
Then the full Hamiltonian will take the form:
\beq
{\cal H}_B = {\cal H} + A_0 \Lambda -
\frac{\xi}{2} E_0^2 + A^k\partial_kE_0 +
\overline{\Pi} \Pi + \overline{C} (-\Delta) C,
\l{7c.3}
\eeq
An additional term with ghost fields is added.  For  the  model  under
consideration, ghosts do not interact with other fields,  they satisfy
the equations:
\beb
\partial_{\mu} \partial^{\mu} C(x) = 0,
\quad
\partial_{\mu} \partial^{\mu} \overline{C}(x) = 0,
\\
\left[C(x),C(y)\right]_+ =0, \quad
\left[\overline{C}(x),\overline{C}(y)\right]_+ =0, \\
\left[C(x),\overline{C}(y)\right]_+ = -i D_0(x-y),
\l{7c.4}
\eeb
while the BRST-cherge is taken to the form:
\beq
\hat{Q} = \int d\sigma^{\mu}
[\partial_{\mu} C E_0 - C \partial_{\mu} E_0].
\l{7c.5}
\eeq
One supposes that ghosts are in vacuum; then condition
\r{7c.1b} is taken to the form \r{6.10}.

Notice that  BRST-approach  allows  us  to   construct   a   covariant
fomulation of  Abelian  Higgs model with spontaneous symmetry breaking
in `t Hooft gauge (cf. \c{31}). The ghosts interact with scalar field,
so that they should be taken into account.

The BRST-approach  is  applicable  to  non-Abelian  theories  as well
\c{26}.

To investigate  semiclassical  gauge  transformations,  we  need  some
properties of integrals of motion.

\subsection{Noether integrals of motion}

Investigate properties of the Noether integrals of motion
$\hat{Q}$ in semiclassical field theory. Commutation relations between
$\hat{Q}$ and $K_X^h$ may be written analogously to
\r{2.5} as
\beq
\hat{Q} K_X^h f = K_X^h \underline{Q}_X f,
\l{7a.1a}
\eeq
semiclassical charge
$\underline{Q}_X$  is expanded in $\sqrt{h}$:
\bez
\underline{Q}_X = Q_X + \sqrt{h} Q_X^{(1)} + ...
\eez
and coincide in the leading order with classical integral of motion
$Q_X$.

Analogously to  properties  of the field operator \r{2.6v},  \r{2.13n}
and \r{2.16}, one obtains the relations
\beb
\underline{Q}_{u_gX}
\underline{U}_g(u_gX \gets X) =
\underline{U}_g(u_gX \gets X)
\underline{Q}_X,
\\
ih \frac{\partial}{\partial \alpha} \underline{Q}_X =
\left[
\underline{Q}_X,\underline{\omega}_X[\frac{\partial       X}{\partial
\alpha}]\right],
\quad
\underline{Q}_{X_2} \underline{V}(X_2\gets X_1) =
\underline{V}(X_2\gets X_1)
\underline{Q}_{X_1}.
\l{7a.2}
\eeb
When the covariant formulation is used, the relations takes the form:
\beq
\underline{Q}_{u_gJ}
\underline{U}_g = \underline{U}_g \underline{Q}_J,
\l{7a.3}
\eeq
\beq
ih \frac{\delta \underline{Q}_J}{\delta J(x)}
= [\underline{\Phi}_R(x|J),\underline{Q}_J].
\l{7a.3a}
\eeq
and
\bez
\underline{W}_{J_2+J_+}
\underline{Q}_{J_2}
\underline{W}_{J_2+J_+}^+
=
\underline{W}_{J_1+J_+}
\underline{Q}_{J_1}
\underline{W}_{J_1+J_+}^+,
\eez
here $supp J_+ > supp J_{1,2}$,  $J_1+J_+\sim 0$,  $J_2+J_+\sim 0$. It
follows  from  \r{7a.3a}  that when one variates $J\to J+\delta J$ the
variation           $\underline{W}_{J+J_+}           \underline{Q}_{J}
\underline{W}_{J+J_+}^+$  will  vanish.  This  means  that  the  third
property \r{7a.2} is a corollary of others.

It happens that integrals of motion corresponding to symmetry
transformations
\bez
\Phi(x) \to \Phi(x) + \overline{\delta} \Phi(x|\Phi),
\eez
have the following form \c{20} in the leading order:
\beq
Q_J = \int dx J(x) \overline{\delta} \Phi
(x|\Phi_R(\cdot|J)).
\l{7a.4a}
\eeq

\subsection{Properties of semiclassical BRST-charge}

Let us discuss specific features of  covariant  formulation  of  gauge
theories. First, introduce semiclassical fermionic fields:
antighost $\ou{C}(x|J)$  and ghost
$\underline{C}_R(x|J)$.

{\bf C6 (BRST)}. {\it
To each classical source  $J(x)$  with  compact  support  one  assigns
operators   of   semiclassical   fields   of   ghosts  and  antighosts
$\underline{C}_R(x|J)$ (Hermitian) and $\ou{C}(x|J)$  (anti-Hermitian)
expanded  in  $\sqrt{h}$.  The  field  $\underline{C}_R(x|J)$  depends
on $J$ at preceeding time moments only;  the following properties  are
satisfied:
\beb
\underline{U}_{g^{-1}} \underline{C}_R(x|u_gJ) \underline{U}_{g}
= \underline{C}_R(w_gx|J),
\quad
\underline{U}_{g^{-1}} \ou{C}(x|u_gJ) \underline{U}_{g}
= \ou{C}(w_gx|J),
\\
ih \frac{\delta \underline{C}_R(x|J)}{\delta J(y)}
= [\underline{\Phi}_R(y|J),\underline{C}_R(x|J)],
\quad x\gsim y,
\\
ih \frac{\delta \ou{C}(x|J)}{\delta J(y)}
= [\underline{\Phi}_R(y|J),\ou{C}(x|J)].
\l{7e.1}
\eeb
The operator    $\underline{W}[\Phi_c]$    satisfies   the   following
commutation relations:
\beb
\underline{W}[\Phi_c] \underline{C}_R(x|0) \underline{W}[\Phi_c] =
\underline{C}_R(x|J_{\Phi_c}),
\quad x\gsim supp \Phi_c,
\\
\underline{W}[\Phi_c] \ou{C}(x|0) \underline{W}[\Phi_c] =
\ou{C}(x|J_{\Phi_c}).
\l{7e.6}
\eeb
}

Take into account the relations imposed on the BRST-charge  and  ghost
number.

{\bf C7 (BRST)}. {\it
To each classical source $J(x)$ with compact support one  assigns  the
operators  of  semiclassical  BRST-charge  $\underline{Q}_J$ and ghost
number $\underline{N}_J$,  expanded in $\sqrt{h}$.  They satisfies the
property:
\beb
\underline{Q}_{u_gJ} \underline{U}_g =
\underline{U}_g \underline{Q}_J,
\quad
ih \frac{\delta \underline{Q}_J}{\delta J(y)}
= \left[\underline{\Phi}_R(y|J), \underline{Q}_J\right],
\\
\underline{N}_{u_gJ} \underline{U}_g =
\underline{U}_g \underline{N}_J,
\quad
ih \frac{\delta \underline{N}_J}{\delta J(y)}
= \left[\underline{\Phi}_R(y|J), \underline{N}_J\right],
\\
\left[ \underline{N}_J, \underline{C}_R(x|J) \right] =
\underline{C}_R(x|J),
\quad
\left[ \underline{N}_J, \ou{C}(x|J) \right] =
\ou{C}(x|J),
\quad
\left[ \underline{N}_J, \underline{\Phi}_R(x|J) \right] =
0.
\\
\left[\underline{N}_J,\underline{Q}_J\right] = \underline{Q}_J,
\quad
\underline{Q}_J \underline{Q}_J = 0,
\quad
\underline{Q}_J^+ = \underline{Q}_J.
\l{7e.2}
\eeb
\beq
\left[\underline{Q}_J, \ou{C}(x|J)\right]_+
= - i \underline{\cal E}_0(x|J).
\l{7e.3}
\eeq
}

Find an explicit form of the BRST-charge in the leading order  of  the
expansion.  Consider  the  scalar  electrodynamics  as an example.  It
follows from \r{7c.5} that
\bez
\underline{Q}^{(1)}_J =
\int d\sigma^{\mu}
[\partial_{\mu} C_R^{(1)} {\cal E}_{0R}^{(1)} -
C_R^{(1)} \partial_{\mu}{\cal E}_{0R}^{(1)}];
\eez
the integral is taken over any space-like surface at the domain
$x> supp J$. Making use of relations
\bey
\partial_{\mu} \partial^{\mu} C_R^{(1)} = 0,
\\
\partial_{\mu} \partial^{\mu} {\cal E}_{0R} =
\partial_{\mu} J^{\mu} -  i  (\sigma^*\Theta_R  -  \Theta_R^*  \sigma),
\eey
take BRST-charge to the form
\beq
Q_J^{(1)} = \int dx
[J^{\mu} \partial_{\mu} C^{(1)}_R +
i  (\sigma^*\Theta_R  -  \Theta_R^*  \sigma) C_R^{(1)}].
\l{7d.2}
\eeq
We see that formula
\r{7d.2} is analogous to relation
\r{7a.4a}  for Noether integral of motion, provided that
$\overline{\delta} \Phi$
is an infinitesimal transformation with anticommuting variables
\beq
\overline{\delta} {\cal A}_{\mu} = \partial_{\mu} C_R^{(1)},
\quad
\overline{\delta} \Theta = i\Theta C_R^{(1)},
\quad
\overline{\delta} \Theta^* = - i\Theta C_R^{(1)},
\l{7d.3}
\eeq
which is analogous to gauge transformation. Property
\r{7d.2} can be viewed as a basis for
the semiclassical perturbation theory for
BRST-charge.

Consider the conditions on the physical states.

{\bf C8   (BRST)}   {\it   Physical   states   satisfying   conditions
$\underline{Q}_J  \overline{f}  = 0$,  $\underline{N}_J \overline{f} =
0$,  have non-negative square of norm.  $(\overline{f},\overline{f}) =
0$ iff $\overline{f} = \underline{Q}_J \overline{g}$. }

Analogously to  electrodynamics,  introduce  equivalence  relation for
sourses and semiclassical states. An analog of property
\r{6.24} will have the following form.

{\bf C9  (BRST)}.  The operator of semiclassical BRST-charge obeys the
condition
\beq
\underline{Q}_{J+\Delta\kappa} = \underline{Q}_J,
\quad
supp \Delta \kappa > supp J.
\l{7d.1}
\eeq

\subsection{Leading order of semiclassical expansion}

Investigate the properties of BRST-charge  in  the  leading  order  of
semiclassical expansion.  Suppose that zero order of expansion for the
BRST-charge, classical action, ghost and antighost fields
$c_0(x)$ and  $\overline{c}_0(x)$  as  $J=0$ is known.  These are free
fields.

Denote
$\Phi^{(1)}_R(x|J) \equiv
(\hat{A}^{\mu}(x),\hat{E}_0(x),\hat{\theta}(x),\hat{\theta}^*(x))$,
$c(x) \equiv C_R^{(1)}(x|J)$.

It follows  from  commutation  relation  \r{7e.1}  that  the operators
$c(x)$      and      $\overline{C}^{(1)}(x|J)$      commute       with
$(\hat{A}^{\mu}(x),\hat{E}_0(x),\hat{\theta}(x),\hat{\theta}^*(x))$.

It follows  from  the  nilpotent  property of the BRST-charge that the
operators $c(x)$ should anti-commute
\bez
[c(x),c(y)]_+ = 0.
\eez
Write down the commutation relation
\r{7e.2} in the second order of perturbation theory.
For electrodynamics, one obtains:
\beb
\left[\hat{A}^{\mu}(x),
\underline{Q}_J^{(2)} \right] = i
\partial^{\mu} c(x),
\quad
\left[
\hat{E}_0(x),
\underline{Q}_J^{(2)} \right] = 0,
\\
\left[\hat{\theta}(x),
\underline{Q}_J^{(2)} \right] = -\Theta_R(y|J) c(y),
\quad
\left[\hat{\theta}^*(x),
\underline{Q}_J^{(2)} \right] = \Theta_R^*(y|J) c(y).
\l{7e.5}
\eeb
Making use of eq.\r{6.30b}, we find that
\bez
\partial_{\mu} \hat{A}^{\mu} + \xi \hat{E}_0 = 0,
\eez
therefore, the ghost field $c(x)$ satisfies the free equation
\bez
\partial_{\mu} \partial^{\mu} c(x) = 0.
\eez
At $x  <  supp  J$,  field  $c(x)$  should  coincide to the free field
$c_0(x)$; therefore, property $c(x) = c_0(x)$ is valid for all $x$.

It follows from property \r{7e.3} in the leading order that
\bez
[c_0(x), \overline{C}^{(1)}(y|J)] = - i D_0(x-y).
\eez
Then
\bez
\overline{C}^{(1)}(y|J) = \overline{c}_0(y),
\eez
since the operator
$\overline{C}^{(1)}(y|J)$
should be linear combination of
$\overline{c}_0$.

It forllows from commutation relations
\r{7e.3}   and  \r{7e.5}  that
\bez
\underline{Q}_J^{(2)} = \int d{\bf x}
[\dot{c}({\bf x}) E_0({\bf x}) -
c({\bf x}) \dot{E}_0({\bf x})].
\eez
Properties \r{7e.6} mean  that  the  operator  $\underline{W}[\Phi_c]$
should commute with ghost and antighosts.

Higher orders are constructed analogously.

\newpage
\pagestyle{empty}

\end{document}